\documentclass[fleqn,usenatbib,useAMS]{mnras}

\usepackage{graphicx}	
\usepackage{amsmath}	
\usepackage{amssymb}	
\usepackage{multicol}        
\usepackage{bm}		
\usepackage{pdflscape}	
\usepackage{hyperref} 

\newcommand{\deriv}[2]{\frac{{\mathrm d} #1}{{\mathrm d} #2}}
\newcommand{\diff}[2]{\frac{{\mathrm D} #1}{{\mathrm D} #2}}
\newcommand{\pder}[2]{\frac{\upartial #1}{\upartial #2}}
\newcommand{\pderN}[3]{\frac{\upartial^{#3} #1}{\upartial #2^{#3}}}
\newcommand{\derivN}[3]{\frac{d^{#3} #1}{d #2^{#3}}}
\newcommand{\pdXnYm}[6]{\frac{\upartial^{#1} #2}{\upartial #3^{#4}\upartial #5^{#6}}}
\newcommand{\vect}[1]{\mathbfit #1}
\newcommand{\grad}{ {\bf \nabla } }
\newcommand{\curl}{ {\bf \nabla} \times}
\newcommand{\tv}{\textit v}

\usepackage[T1]{fontenc}
\usepackage{ae,aecompl}

\title[ \LaTeX\ ]{{Dynamics of fast and slow magnetoacoustic waves in plasma slabs with thermal misbalance}}

\author[D. V. Agapova et al.]{D. V. Agapova$^{1,2}$, S. A. Belov$^{1,2}$, N. E. Molevich$^{1,2}$, D. I. Zavershinskii$^{1,2}$%
\thanks{Contact e-mail: \href{mailto:dimanzav@mail.ru}{dimanzav@mail.ru}}%
\thanks{Present address: 34, Moskovskoye shosse, Samara, 443086, Russia}%
\\
$^{1}$Department of Physics, Samara National Research University, Moscovskoe sh. 34, Samara, 443086, Russia\\
$^{2}$Department of Theoretical Physics, Lebedev Physical Institute, Novo-Sadovaya st. 221, Samara, 443011, Russia}

\date{Last updated 2020 June 10; in original form 2013 September 5}

\pubyear{2021}

\begin{document}
\label{firstpage}
\pagerange{\pageref{firstpage}--\pageref{lastpage}}
\maketitle 

\begin{abstract}

Non-uniformity of the solar atmosphere along with the presence of non-adiabatic processes such as radiation cooling and unspecified heating can significantly affect the dynamics and properties of magnetoacoustic (MA) waves. To address the co-influence of these factors on the dispersion properties of MA waves, we considered a single magnetic slab composed of the thermally active plasma. Using the perturbation theory, we obtained a differential equation that determines the dynamics of the two-dimensional perturbations. Applying the assumption of strong magnetic structuring, we derived the dispersion relations for the sausage and kink MA modes. The numerical solution of the dispersion relations for the coronal conditions was performed to investigate the interplay between the non-uniformity and the thermal misbalance. For the heating scenario considered, it was obtained that the phase speed of both the sausage and kink slow MA waves is highly affected by the thermal misbalance in the long wavelength limit. The obtained characteristic timescales of the slow waves dissipation coincide with the periods of waves observed in the corona. Simultaneously, the phase speed of the fast waves is not affected by the thermal misbalance. The geometry of the magnetic structure still remains the main dispersion mechanism for the fast waves. Our estimation reveals that dissipation of the fast waves is weaker than dissipation of the slow waves in the coronal conditions. The obtained results are of importance for using the magnetoacoustic waves not only as a tool for estimating plasma parameters, but also as a tool for estimating the non-adiabatic processes.

\end{abstract}




\begingroup
\let\clearpage\relax
\endgroup
\newpage

\section{Introduction}
	\label{s:Introduction} 
Non-uniformity of the solar atmosphere is largely associated with a strong magnetic field that provides for formation and long existence of various plasma structures like coronal loops, plumes and prominences. 
In fact, these structures are perfect waveguides for the so-called magnetohydrodynamic (MHD) modes \citep[see e.g.][for recent reviews]{Nakariakov2020,2021SSRv..217...76B}. 

Due to the magnetic field, there are two pressure contributions into the plasma: the gasdynamic and magnetic pressure. When these forces are acting together and in opposition, they give rise to rapid and slow compression perturbations that are known as the fast and slow magnetoacoustic (MA) waves, respectively \citep[see][for details]{roberts_2019,2014masu.book.....P}. In the magnetically structured plasma, these waves may have some additional properties. In particular, MA perturbations have possibilities to be trapped or to leak out of the magnetic structure (trapped/leaky modes), to evolve inside the structure or only near its boundary (body/surface modes), to have a different axial symmetry (kink/sausage/fluting modes), etc. As a result, there are a wide variety of wave types corresponding to a different evolution way of compression perturbation.

Nowadays, with the help of groundbased and orbital instruments, there is a plenty of observational data (intensity variations, Doppler shifts, etc) that can be associated with various types of MA modes. In particular, a recent review concerning the slow MA waves can be found in  \citep{2021SSRv..217...34W}. An extensive discussion on the fast kink modes is presented in  \citep{2021SSRv..217...73N}. Some observations and modelling of fast sausage waves are discussed in a recent review  by \cite{2020SSRv..216..136L}.

The combination of the MHD theory and observational data allows us not only to associate the observations with a particular mode type but also to use waves as a diagnostic tool for in situ conditions in the solar atmosphere: the local Alfv\'en speed \citep{2017ApJ...837L..11C}, transfer coefficients \citep{2015ApJ...811L..13W}, adiabatic index \citep{2021SoPh..296..105P}, magnetic field strength \citep{Jess2016} etc. The idea that waves could be used to unravel seismological information about the solar atmosphere has become fundamental for the MHD- and coronal seismology – a branch of solar physics probing the parameters of the upper solar layers by MHD waves and oscillations. The pioneering works in this scientific field are the studies by  \citet{1975IGAFS..37....3Z, 1982SvAL....8..132Z} and \citet{1983SoPh...88..179E}, devoted to MHD waves propagation in a magnetic cylinder that is in magnetostatic balance with the surrounding ideal plasma and is composed of ideal plasma, as well.

{In reality}, the coronal plasma is non-ideal, since the solar atmosphere is subject to such non-adiabatic processes as radiation cooling and unspecified heating. Thus, for a magnetic structure to exist for a long time, not only a mechanical balance but also some equilibrium between heating/cooling processes have to be stated. In turn, coronal radiative cooling is known as a function of density and temperature  \citep{2021ApJ...909...38D}. Unspecified heating is also often modelled as a function of plasma parameters \citep{1978ApJ...220..643R,2006A&A...460..573C}. Due to this fact, some compression perturbation in the medium (e.g. MA wave, or entropy/thermal wave) can destroy the balance between these processes and the enhanced or suppressed heating/cooling will subsequently affect the compression waves back. As a result, some feedback between compression waves and non-adiabatic processes will take place. Such effect is known as thermal misbalance or heating/cooling misbalance.

A recent review concerning the problem of thermal misbalance in the solar atmosphere can be found in \citep{10.1088/1361-6587/ac36a5}.  It was shown by \cite{Zavershinskii2019} and \cite{2019A&A...628A.133K} that the thermal misbalance can significantly affect the dispersion properties of slow waves in the solar corona. This leads to the dependence of the phase speed and growth/decay rate on the wave period and can cause amplification of waves or their additional damping.  As a result, the heating/cooling  misbalance may lead to the formation of quasi-periodic patterns \citep[see, e.g.,][]{Zavershinskii2019} at the linear stage, or even to a sequence of self-sustaining shock pulses \citep[see, e.g.,][]{2020PhRvE.101d3204Z,2010PhPl...17c2107C} at the non-linear stage. It was also shown that heating/cooling processes are responsible for the additional phase shift between perturbations of various plasma parameters (density, temperature, etc.) and define the distribution of energy in and between eigenmodes, e.g. slow and entropy modes. Moreover, depending on the heating/cooling mechanisms, the efficiency of the feedback between plasma and eigenmodes may vary significantly leading to a wide variety of evolution scenarios for some initial perturbation \citep[see][for details]{2021SoPh..296...96Z}.  

\cite{Duckenfield2020effect} found that the damping times of slow waves due to thermal misbalance are of the order of $10$\,--\,$100$ minutes, which coincides with the wave periods and damping times observed. The observed temperature dependence of the polytropic index \citep[see, e.g.,][]{2018ApJ...868..149K,2011ApJ...727L..32V} can  be attributed to the thermal misbalance effect, as in such medium it become a function of the heating/cooling rates \citep[see][for detail]{Zavershinskii2019}. In addition, the role of thermal misbalance in estimating the phase shifts is found to be significant for high-density and low-temperature loops.  \cite{2021SoPh..296..105P} showed that variation of heating mechanism can lead to around a five-fold increase in the phase difference.

The influence of both magnetic structuring and thermal misbalance on the dispersion properties of slow sausage waves was investigated by \citet{Belov2021} using the thin flux tube approximation \citep{Zhugzhda96}. It was shown that the frequency dependence of the phase speed is affected by two features: geometric dispersion and dispersion caused by the thermal misbalance. In contrast to the phase speed, the wave decrement is primarily affected by thermal misbalance only. Moreover, it was demonstrated that neglecting the thermal misbalance may be the reason for substantial divergence between the seismological and spectrometric estimations of the plasma parameters.

The analysis conducted by  \citet{Belov2021} is restricted by the long wavelength limit and, moreover, does not account for the external medium and, thus, does not describe the fast waves that are sensitive to its parameters.   Without  the wavelength restrictions, the thermal misbalance was analyzed in the slab geometry by \cite{1991SoPh..131...79V}. However, they focused on the instability of the entropy/thermal mode.  

In this paper, we will focus on the properties of fast and slow MA waves inside a magnetic slab composed of thermally active plasma without restriction on the wavelength. In Section \ref{s:Model}, we discuss the basic equations, used assumptions and introduced  characteristic scales. Further, in Section \ref{s:Equation}, we obtain the differential equation that determines the dynamics of two-dimensional perturbations in an inhomogeneous magnetically structured medium with the thermal misbalance. Section \ref{s:Dispersion} is devoted to the dispersion relation for fast and slow kink/sausage modes in the magnetic slab under the assumption of strong magnetic structuring. Further, in Section \ref{s:Corona}, we apply the obtained results to the solar corona conditions. The discussion and conclusions are presented in Section \ref{s:Discussion}.

\section{Model}
	\label{s:Model} 

Let us consider a fully ionized plasma where the processes of heating and radiative cooling take place. The dynamics of waves and oscillations in such plasma can be described by the system of MHD equations with an additional term corresponding to the non-adiabatic processes in the right hand side of the energy equation: 
\begin{equation}
	\pder{\vect{B}}{t}=\curl\left(\vect{v}\times\vect{B}\right),
	\label{Induction}
\end{equation}
\begin{equation}
	\grad\cdot\vect{B}=0 \,,
	\label{Div}
\end{equation}
\begin{equation}
	\rho\diff{\vect{v}}{t}=-\grad P-\dfrac{1}{4\upi}{\vect{B}}\times\left(\curl\vect{B}\right),
	\label{Motion}
\end{equation}
\begin{equation}
	\pder{\rho}{t}+\grad\cdot\left(\rho{\vect{v}}\right)=0 \,,
	\label{Cont}
\end{equation}
\begin{equation}
	\frac{{\rho}^{\gamma}}{\gamma-1}\diff{}{t}\left(\frac{P}{{\rho}^{\gamma}}\right)=-\rho Q(\rho,T)\,,
	\label{Heat}
\end{equation}
\begin{equation}
	P=\frac{k_\mathrm{B}}{m}\rho T\,.
	\label{State}
\end{equation}
Here $\rho$, $T$, and $P$ respectively represent the density, temperature, and pressure of the plasma, while $\vect{v}$ and $\vect{B}$ are the vectors of the plasma velocity and magnetic field. The Boltzmann constant and the mean mass per volume, are respectively shown by $k_\mathrm{B}$, and $m$. {The adiabatic index is  $\gamma = C_P/C_V = 5/3 $, where  $C_V=3k_\mathrm{B}/2m$ and $C_P=C_V+k_\mathrm{B}/m$ are the specific heat capacities at constant volume and pressure, respectively.} In addition, $D/{D\textit{t}}=\upartial/\upartial t+\mathbfit{v}\cdot\nabla$ stands for the convective derivative. We use the heat-loss function $Q\!\left(\rho, T\right)\!=\!L\!\left(\rho, T\right)-H\!\left(\rho, T\right)$, which is the difference between radiative cooling $L\!\left(\rho, T\right)$ and heating $H\!\left(\rho, T\right)$. The stationary state of the medium implies that non-adiabatic processes balance each other $\!L\!\left(\rho_0, T_0\right) = H\!\left(\rho_0, T_0\right)$, or $Q\left(\rho_0, T_0\right)= 0 $.

As we have mentioned above, the compression perturbation can disturb the thermal balance and thus some interaction/feedback between the plasma and this perturbation will take place. It turned out that the intensity of this interaction and its consequences for waves are highly sensitive to the scale of  perturbation \citep[see][]{Zavershinskii2019}, i.e. its frequency or wavelength. 

This particularly concerns the dependence of the phase velocity  on the frequency $ c_{ph} = c_{ph}(\omega) $  caused by the thermal activity of the plasma. Using the characteristic timescales:
\begin{equation}
	\tau_V={C_V}/{Q_{0T}}, \qquad \tau_P={C_P}T_0/\left({Q_{0T}} T_0-{Q_{0\rho}} \rho_0\right), 
	\label{characteristic_times}
\end{equation}
one may introduce the ranges of weak ($\omega\, $$\left|\tau_{V,P}\right|\gg1$) and strong ($\omega\,$$\left|\tau_{V,P}\right|\ll1$) impact of the thermal misbalance on the  wave phase speed.  Here, $\omega$ is the wave frequency.  $Q_{0T}=\left.\upartial{Q}/\upartial{T}\right|_{\rho_0, T_0},Q_{0\rho}=\left.\upartial{Q}/\upartial{\rho}\right|_{\rho_0, T_0}$. It was shown by  \cite{Molevich88} and \cite{Zavershinskii2019} that the phase velocity $c_{ph}(\omega) $  of the slow modes in the homogeneous thermally active  plasma varies from $c_\mathrm{S}$ to $c_\mathrm{SQ}$, where
\begin{equation}
	c_\mathrm{S}=\sqrt{\gamma \frac{ k_\mathrm{B}T_0}{m}},\qquad c_\mathrm{SQ} =\sqrt{\frac{\tau_V}{\tau_P } \gamma \frac{k_\mathrm{B}T_0}{m}}. 
	\label{characteristic_speeds}
\end{equation}
The harmonics that are weakly affected by the thermal misbalance ($\omega\, $$\left|\tau_{V,P}\right|\gg1$) are propagating  with $	c_\mathrm{S}$, which is the standard value for plasma without thermal misbalance, while in the range ($\omega\,$$\left|\tau_{V,P}\right|\ll1$), the heating and cooling processes completely determine the speed of slow waves $ c_\mathrm{SQ} $.

\begin{figure}
	\includegraphics[width=\columnwidth]{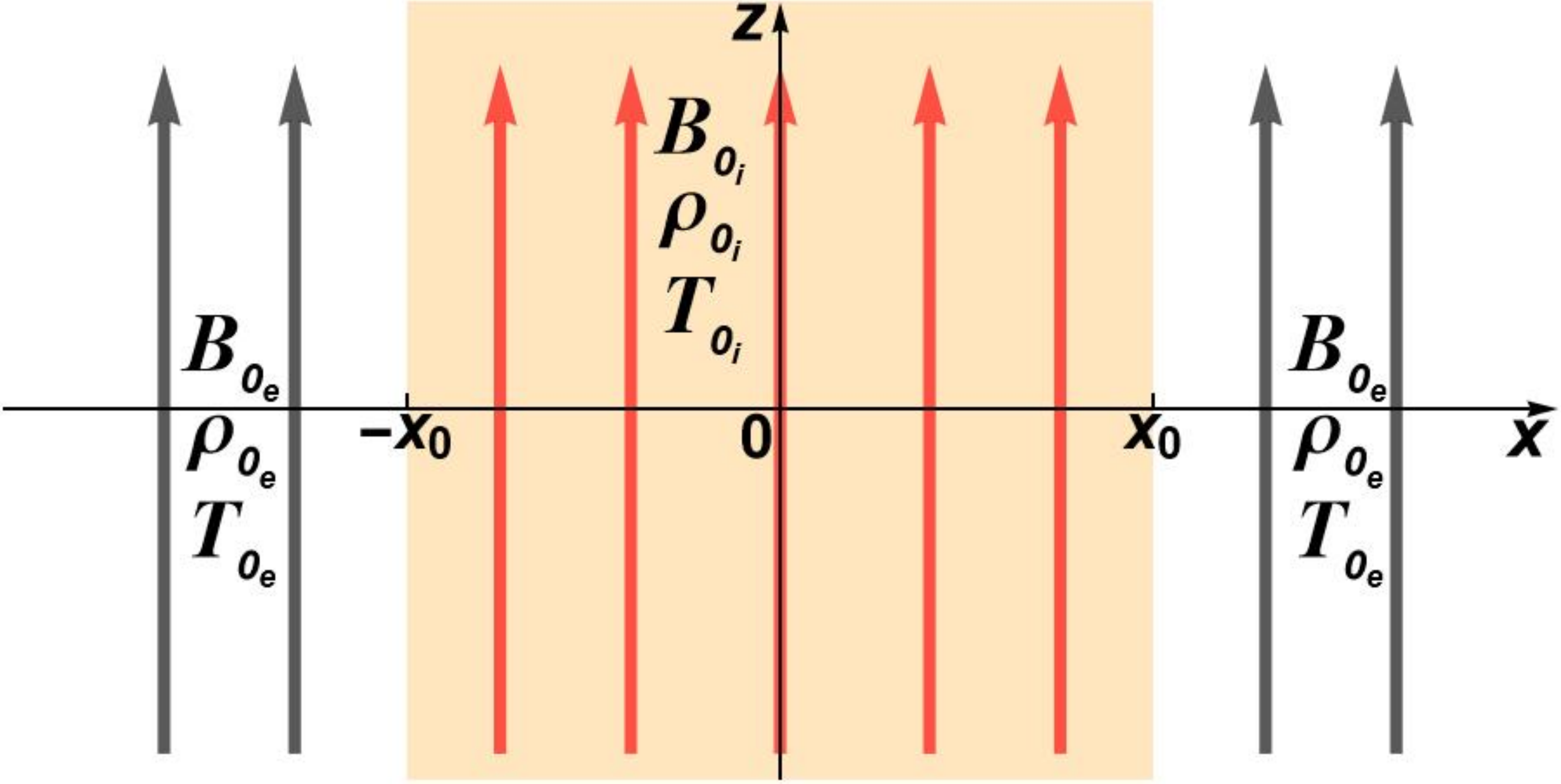}
	\caption{ The configuration of a magnetic slab in the thermally active plasma. The magnetic field lines $B_i$ directed along the $\mathit{z}$-axis are confined to the slab $ -x_0 \le x \le x_0 $ and are surrounded by magnetic field  $B_e$ of the external plasma. The internal and external plasma density, pressure and temperature are $\rho_i, P_i, T_i$ and $\rho_e, P_e, T_e$, respectively. The balance of total pressure (\ref{Press}) is maintained across the slab boundaries $x=\pm x_0$. It is also assumed that the plasma inside and outside the slab is in the thermal balance for a steady state, i.e  $ L\left(\rho_{0_{i}}, T_{0_{i}}\right) = H\left(\rho_{0_{i}}, T_{0_{i}}\right) $ and $ L\left(\rho_{0_{e}}, T_{0_{e}}\right) = H\left(\rho_{0_{e}}, T_{0_{e}}\right) $. }
	\label{fig_Slab}
\end{figure}

Speaking of the problem geometry, the two basic models generally used to analyze waves in the magnetically structured plasma, namely, magnetic cylinder/tube \citep[see][]{1983SoPh...88..179E} and magnetic slab \citep[see][]{Edwin1982}. The modes in such geometrical objects as the slab and
the tube are closely related in many aspects. However, any treatment of waves in a tube involves introduction of Bessel or Hankel functions. Due to the fact that this complication is avoided in the slab, such geometry is often used both for analytical and forward numerical modelling. In this work, we will also adhere to the geometry of the magnetic slab (see Fig. \ref{fig_Slab}).

\section{Small perturbations in magnetically structured plasma}
	\label{s:Equation} 
Next, we consider an unperturbed magnetic field $\mathbfit{B}_\textbfit{0}\textbfit{(x)}$ directed along the \textit{z}-axis and varying in the perpendicular direction along the \textit{x}-axis. To satisfy the stationary conditions, the unperturbed pressure and density should also vary along the \textit{x}-axis: $P_0(x)$, $\rho_0(x)$. Let us investigate small amplitude perturbations using  the standard perturbation theory method. In this case, we can use the following replacements for equations~(\ref{Induction}) -- (\ref{State}):
\begin{align}\label{ent_av_inc_lim}
	&	\rho=\rho_0(x)+\rho_1, \, P=P_0(x)+P_1, T=T_0(x)+T_1, \\
	& \mathbfit{B}=\mathbfit{B}_\textbfit{0}\textbfit{(x)}+\mathbfit{B}_\textbfit{1}, \, \mathbfit{v}=\mathbfit{v}_\textbfit{1}.\nonumber
		\label{Сhange}
\end{align}
Hereinafter, index ''0'' means the unperturbed state of plasma, numerical index ''1'' means the perturbation of stationary profile, alphabetic subscripts (''x'', ''y'', ''z'') mean the corresponding component of the vector. Investigation of the small amplitude perturbations means that
\begin{multline}
	\frac{\rho_1}{\rho_0}\thicksim\frac{P_1}{P_0}\thicksim\frac{T_1}{T_0}\thicksim\frac{{B_1}_x}{B_0}\thicksim\frac{{B_1}_y}{B_0}\thicksim\frac{{B_1}_z}{B_0}\thicksim\\
	\thicksim\frac{{v_1}_x}{c_\mathrm{S}}\thicksim\frac{{v_1}_y}{c_\mathrm{S}}\thicksim\frac{{v_1}_z}{c_\mathrm{S}}\thicksim\varepsilon\ll1.
\end{multline}

After a number of transformations, the system of equations~(\ref{Induction}) -- (\ref{State}) can be reduced to the form:
\begin{multline}
	\pder{}{t}\left(\pder{P_T}{t}-\rho_0c_\mathrm{A}^2\pder{{\tv_1}_z}{z}+\rho_0\left(c_\mathrm{A}^2+c_\mathrm{S}^2\right)\!{\Theta}\right)\!\tau_V +\\
	+\pder{P_T}{t}\! -\rho_0c_\mathrm{A}^2\pder{{\tv_1}_z}{z}+\!\rho_0\left(c_\mathrm{A}^2+c_\mathrm{SQ}^2\right){\Theta}+\\
	+{\tv_1}_x\left(\deriv{}{x}\left(\frac{B^2_0}{8\upi}\right)+c_\mathrm{SQ}^2\deriv{\rho_0}{x}\right)=0,
	\label{Connect}
\end{multline}
\begin{equation}
	\pderN{{\tv_1}_x}{t}{2}=-\frac{1}{\rho_0}\pdXnYm{2}{P_T}{x}{}{t}{}+c_\mathrm{A}^2\pderN{{\tv_1}_x}{z}{2},
	\label{x-comp}
\end{equation}
\begin{equation}
	\pderN{{\tv_1}_y}{t}{2}=-\frac{1}{\rho_0}\pdXnYm{2}{P_T}{y}{}{t}{}+c_\mathrm{A}^2\pderN{{\tv_1}_y}{z}{2},
	\label{y-comp}
\end{equation}
\begin{equation}
	\pderN{{\tv_1}_z}{t}{2}=-\frac{1}{\rho_0}\pdXnYm{2}{P_T}{z}{}{t}{}+c_\mathrm{A}^2\left(\pderN{{\tv_1}_z}{z}{2}-\pder{\Theta}{z}\right),
	\label{z-comp}
\end{equation}
where $P_T\!=\!P_1+({B_0}/{4\upi}){B_1}_z$ is the total pressure (gas-dynamic plus magnetic) perturbation; $\Theta=\nabla\cdot\mathbfit{v}_1$; $c_\mathrm{A}\!=\!\sqrt{B_0/{4\upi\rho_0}}$ is the Alfvén velocity.

Next, let us use the following Fourier substitution: 
\begin{eqnarray}
	\tv_{1x}={\tilde{\tv}}_{1x}\left(x\right)e^{i\left(\omega t+k_yy+k_zz\right)},\\
	\tv_{1y}={\tilde{\tv}}_{1y}\left(x\right)e^{i\left(\omega t+k_yy+k_zz\right)},\\
	\tv_{1z}={\tilde{\tv}}_{1z}\left(x\right)e^{i\left(\omega t+k_yy+k_zz\right)}, \\
	 P_T={\tilde{P}}_T\left(x\right)e^{i\left(\omega t+k_yy+k_zz\right)},
\end{eqnarray}
where ${\tilde{\tv}}_{1x}$, ${\tilde{\tv}}_{1y}$, ${\tilde{\tv}}_{1z}$, ${\tilde{P}}_T$ -- amplitude components of the velocity vector and total pressure perturbation; $k_y$, $k_z$ are the wave numbers in $\mathit{y}$- and $\mathit{z}$-directions, respectively.

 Using this substitution and combining equations~(\ref{Connect}) -- (\ref{z-comp}), we can obtain the equation for the amplitude of total pressure perturbation:
\begin{multline}
	{\tilde{P}}_T=\left[\frac{\rho_0}{\omega^2}\left(i\omega\tau_VA^2+A^2_{\mathrm{Q}}\right)\deriv{\tilde{\tv}_{1x}}{x} \right. -\\
	\left.\qquad-\left(\deriv{}{x}\left(\frac{B^2_0}{8\upi}\right)+c_\mathrm{SQ}^2\deriv{\rho_0}{x}\right){\tilde{\tv}}_{1x}\right] \times \\
	 \times 	\frac{i\omega(k_z^2c_\mathrm{A}^2-\omega^2)}{\left(A^2_\mathrm{Q}\left(m^2_\mathrm{Q}+k_y^2\right)+i\omega\tau_VA^2\left(m^2+k_y^2\right)\right)},   
	\label{Pressure}
\end{multline}

where the following notations are introduced:
\begin{eqnarray}
m^2=\frac{\left(k^2_zc^2_\mathrm{A}-\omega^2\right)\left(k^2_zc^2_\mathrm{S}-\omega^2\right)}{\left(c^2_\mathrm{A}+c^2_\mathrm{S}\right)\left(k^2_zc^2_\mathrm{T}-\omega^2\right)}\!,\nonumber\\ m^2_\mathrm{Q}=\frac{\left(k^2_zc^2_\mathrm{A}-\omega^2\right)\left(k^2_zc^2_\mathrm{SQ}-\omega^2\right)}{\left(c^2_\mathrm{A}+c^2_\mathrm{SQ}\right)\left(k^2_zc^2_\mathrm{TQ}-\omega^2\right)},\nonumber\\
A^2=\left(c^2_\mathrm{A}+c^2_\mathrm{S}\right)\left(k^2_zc^2_\mathrm{T}-\omega^2\right)\!,\nonumber\\
A^2_\mathrm{Q}=\left(c^2_\mathrm{A}+c^2_\mathrm{SQ}\right)\left(k^2_zc^2_\mathrm{TQ}-\omega^2\right)\!.\nonumber
\end{eqnarray}
To describe the properties of MHD waves in the thermally active plasma,  it is useful to introduce two additional characteristic values of the phase speed:
\begin{equation}
	c_\mathrm{T}=\sqrt{ \frac{{c^2_\mathrm{S}c^2_\mathrm{A}}}{\left(c^2_\mathrm{A}+c^2_\mathrm{S}\right)}},\qquad c_\mathrm{TQ} =\sqrt{\frac{c^2_\mathrm{A}c^2_\mathrm{SQ}}{\left(c^2_\mathrm{A}+c^2_\mathrm{SQ}\right)}}. 
	\label{characteristic_tube_speeds}
\end{equation}
The value $c_\mathrm{T}$ is a well-known tube speed in the ideal plasma generally associated with slow waves. This speed is a result of geometry dispersion caused by the magnetic structuring only. However, the combination of geometry dispersion and the dispersion caused by the thermal misbalance leads to the emergence of the modified tube-speed  $c_\mathrm{TQ}$, which was introduced by \cite{Belov2021}. As we will show further, the $c_\mathrm{TQ}$ speed is the low-frequency/long-wavelength limit value of the slow wave phase velocity for either kink or sausage modes.


Finally, combining  the Fourier transformation of equation~(\ref{x-comp}) with  equation~(\ref{Pressure}) differentiated by the \textit{x}-coordinate yields:
\begin{multline}
	\deriv{}{x}\left[\frac{\rho_0(k_z^2c_\mathrm{A}^2-\omega^2)\left(i\omega\tau_VA^2+A^2_{\mathrm{Q}}\right)}{\left(A^2_{\mathrm{Q}}\left(m^2_{\mathrm{Q}}+k_y^2\right)+i\omega\tau_VA^2\left(m^2+k_y^2\right)\right)}\deriv{\tilde{\tv}_{1x}}{x}\right. -\\
	\left.-\frac{\omega^2(k_z^2c_\mathrm{A}^2-\omega^2){\tilde{\tv}}_{1x}}{\left(A^2_{\mathrm{Q}}\left(m^2_{\mathrm{Q}}+k_y^2\right)+i\omega\tau_VA^2\left(m^2+k_y^2\right)\right)} \right. \times \\
    \left.  \times 	\left(\deriv{}{x}\left(\!\frac{B^2_0}{8\upi}\right)+c_\mathrm{SQ}^2\deriv{\rho_0}{x}\right)\right]	-\rho_0\left(c_\mathrm{A}^2k_z^2-\omega^2\right){\tilde{\tv}}_{1x}=0.
	\label{Disp}
\end{multline}
The resulting equation (\ref{Disp}) determines the dynamics of two-dimensional perturbations in an inhomogeneous magnetically structured medium with the thermal misbalance.

\section{Dispersion relation}
	\label{s:Dispersion}
For the general case of magnetic structuring (i.e., when $B_0\!\left(x\right)$ is an arbitrary function), a solution of the resulting equation~(\ref{Disp}) can be obtained only numerically. Therefore, in what follows,  we will consider a simpler case of strong magnetic structuring (step-function profiles of density and magnetic field strength). 

Let us analyze a magnetic slab with a width of $2x_0$ and a magnetic field strength $B_i$ inside it surrounded by a plasma with  field $B_e$ (see  Fig.~\ref{fig_Slab}). We consider that the magnetic field is given by the following relationship:
\begin{equation}
	B_0\!\left(x\right)= 
	\begin{cases}	
		B_i,\ \ \left|x\right|\le x_0, \\ 
		B_e,\ \ \left|x\right|>x_0.
	\end{cases}	
	\label{System}
\end{equation}
 Hereinafter, indices ''i'' and ''e'' correspond to the parameters inside and outside the slab, respectively.
 
 To satisfy the condition of mechanical equilibrium, the total pressure must be continuous at the boundaries of the magnetic slab (at $x=\pm x_0 $):
\begin{equation}
	P_e+\frac{B^2_e}{8\upi}=P_i+\frac{B^2_i}{8\upi}.
	\label{Press}
\end{equation}

In addition, for the stationary state, the thermal equilibrium in the internal and external plasma should be established. This implies  that
\begin{equation}
  Q\left(\rho_{0_{i}}, T_{0_{i}}\right) = 0,\qquad Q\left(\rho_{0_{e}}, T_{0_{e}}\right) =0. 
	\label{thermal_balance}
\end{equation}

Now, let us analyze the two-dimensional perturbations and neglect the dependence on the \textit{y}-coordinate. Then, the wavenumber in the \textit{y}-direction $k_y$  should be set to zero. For such perturbations, the transverse component of the velocity vector is given as $\tv_x={\hat{\tv}}_x\left(x\right)e^{i\left(\omega t+k_z z\right)}$, where ${\hat{\tv}}_x\left(x\right)$ is the amplitude of the disturbance. We will study the propagation of waves emerging inside the slab and vanishing outside it (i.e., $ \tv_x \rightarrow0$  at $x\rightarrow\pm\infty$). 

In the case considered, the differential equation for the velocity perturbation ~(\ref{Disp})  can be rewritten in the following forms:
\begin{eqnarray} 
	\derivN{{\hat{\tv}}_{1x}}{x}{2}-k^2_{x_i}{\hat{\tv}}_{1x}=0, \quad  \left|x\right|\le x_0, \nonumber	\\ 
	\derivN{{\hat{\tv}}_{1x}}{x}{2}-k^2_{x_e}{\hat{\tv}}_{1x}=0, \quad  \left|x\right|>x_0,         
	\label{velocity}
\end{eqnarray}

where, to simplify the formula, we introduce $k^2_{x_{i,e}}$ as follows:
	$$k^2_{x_{i,e}}=\frac{\left(A^2_{\mathrm{Q_{i,e}}}\!m^2_{\mathrm{Q_{i,e}}}+i\omega{\tau_V}_{i,e}   A^2_{i,e }m^2_{i,e}\right)}{\left(A^2_{\mathrm{Q_{i,e}}}+i\omega{\tau_V}_{i,e}A^2_{i,e}\right)}.$$
Here, we should mention that  in the thermally active plasma $k_x$ becomes complex. This is different from the ideal plasma case \citep[see][]{Edwin1982},  where the quantity  $k_x$ (or $m$ in the author's original notation) can be either real or pure imaginary depending on the sign of $k_x^2$. In the ideal plasma, the case of $k_x^2>0$ is associated with the surface modes, and the case of $k_x^2<0$ is associated with the body modes. However, due to the $k_x$ complexity, distinguishing between surface and body modes become less clear in the non-adiabatic plasma as the modes are partly surface and partly body.

Nevertheless, some distinguishing can be conducted by the analysis of the real part $Re\left(k_x^2\right)$.  Then, if we obtain that the real part is positively defined $Re\left(k_x^2\right)>0$, we can conclude that $Re\left(k_x\right)>Im\left(k_x\right)$. This means that the wave under consideration is localized near the boundary of the slab and thus can be considered as the ‘rather surface than body’ mode. In the opposite case of $Re\left(k_x^2\right)<0$, the analogy would suggest that  $Re\left(k_x\right)<Im\left(k_x\right)$. Thus, such waves can be called as ''rather body  than surface''. In what follows, we omit the detailed description and call the waves simply surface and body modes implying  ''rather surface than body'' and ''rather body  than surface'', respectively.

Further, we will consider that $Re(k_x)>0$. Thus, solving equations (\ref{velocity}) and taking into account that  ${\hat{v}}_{1x}\!\rightarrow\!0$ for $\left|x\right|\!\rightarrow\!\infty$, which, according to the above mentioned, implies that $Re\left(k^2_{x_{e}}\right)>0$  for $\left|x\right|\!\rightarrow\!\infty$ (depending on the current conditions, inside the slab, the sign of  $Re\left(k^2_{x_{i}}\right)$ can be arbitrary), we obtain
\begin{equation}
	{\hat{v}}_{1x}=\left\{ 
	\begin{array}{lr}
		\alpha_i{\cosh  \left(k_{x_i}x\right)\ }+\beta_i{\sinh  \left(k_{x_i}x\right)\ }, & \left|x\right|\le x_0; \\ 
		\beta_e \exp \left({k_{x_e}\left(x+x_0\right)} \right)	, &  x<{-x}_0;\\
		\alpha_e \exp \left({-k_{x_e}\left(x-x_0\right)}\right), & x>x_0. 
	\end{array}
	\right. 
	\label{vx_solution}
\end{equation}

Solution (\ref{vx_solution}) depends on four unknown constants $\alpha_i$, $\beta_i$, $\alpha_e$ and $\beta_e$. {In order to describe the dispersion properties of waves, these quantities should be determined (up to a multiplicative constants).} To do this, we first need to choose the symmetry type for the mode of interest. After this step is done, to obtain  the dispersion relation for the chosen mode, we should join the solutions inside and outside the slab by applying boundary conditions.

As we have mentioned earlier, there are two well-known symmetry types for waves in the magnetic slab.  These types are the so-called sausage and kink waves corresponding to the symmetric and anti-symmetric perturbations of the slab boundaries, respectively. Then, the choice of the sausage or kink mode requires to set $\alpha_i=0$ or $\beta_i=0$, respectively.

To join the solutions at boundaries, we have two conditions, namely, the  continuity of velocity component  $v_x$ and the continuity of  total pressure $P_T$. The equations for the constants corresponding to the first condition can be easily written using equation (\ref{vx_solution}). To write the equations corresponding to the second condition, one can use the expression for the total pressure perturbation ${\hat{P}}_T$. Using equation (\ref{Pressure}), it can be written in the general form  as: 
 \begin{equation} 
 	{\hat{P}}_T=i\frac{\rho_{0_{i,e}}}{\omega}\frac{(k^2_zc_\mathrm{A_{i,e}}^2\!-\omega^2)}{k^2_{x_{i,e}}}\deriv{{\hat{\tv}}_{1x}}{x}.
 	\label{P_amp}
 \end{equation} 

The substitution of solutions (\ref{vx_solution}) in equation  (\ref{P_amp}) allows us to complete  the system of equations for constants  $\alpha_i$, $\beta_i$, $\alpha_e$ and $\beta_e$, and thus to determine  the dispersion relations. 
 
Finally, the dispersion relations for the full set (fast/slow and body/surface) of sausage/kink magnetoacoustic waves in the magnetic slab composed of the thermally active plasma are:
\begin{equation}
	(k^2_z{c^2_\mathrm{A}}_{i}\!-\omega^2) \frac{k_{x_e}}{k_{x_i}} \!
	=\!-\frac{\rho_{0_e}}{\rho_{0_i}}\left(k^2_z{c^2_\mathrm{A}}_{e}\!-\omega^2\right)\!\!\left( 
	\begin{array}{c}
		{\rm\!coth\ (k_{x_i}x_0)} \\ 
		{\rm\!tanh\ (k_{x_i}x_0)} 
	\end{array}
	\right)\!.
	\label{Solution}
\end{equation}
Here, we use hyperbolic functions $ \rm\!Coth\! $  and $\rm\!Tanh\!$ for kink and sausage modes, respectively. Further, we will apply these equations to calculate the dependencies of phase velocities and increment/decrement of MA waves on wavenumbers in the solar atmosphere conditions. It should be mentioned that in the absence  of thermal misbalance (${\tau_V}_{i,e}\!\rightarrow\!0$), the equations~(\ref{Solution}) reduce to those  obtained  for ideal plasma by \cite{Edwin1982}.

\section{Application to solar corona}
\label{s:Corona}
\subsection{Heating and cooling model}
\begin{table*}
	\centering
	\caption{Slab parameters used for calculations.}
	\label{parameters}
	\begin{tabular}{lcccr} 
		\hline
		Parameter & & Value  & \\
		& ''Warm'' loop & ''Hot'' loop & ''Cool'' loop\\
		\hline
		Temperature inside the slab $\left(T_{0_{i}}\right)$              & $1$\,MK               & $6$\,MK   & $0.6$\,MK \\
		Temperature outside the slab $\left(T_{0_{e}}\right)$             & $1$\,MK               & $6$\,MK   & $0.6$\,MK \\
		Number density inside the slab $\left(n_{0_{i}}\right)$           & $10^{10}$\,cm$^{-3}$  & $10^{11}$\,cm$^{-3}$ &   $10^{9}$\,cm$^{-3}$        \\
		Density contrast $\left(n_{0_{i}}/n_{0_{e}}\right)$               & 5                     & 10         & 2 \\
		Magnetic field strength inside the slab $\left(B_{0_{i}}\right)$  & $10$\,G               & $50$\,G    & $5$\,G  \\
		Slab half-width $\left(x_0\right)$                                & $1$\,Mm               & $1$\,Mm    &$1$\,Mm \\				
		\hline
	\end{tabular}
\end{table*}

Since the obtained equations (\ref{Solution}) are transcendental, we will solve them numerically to highlight the main influence of the thermal misbalance on the dispersion properties of magnetoacoustic waves in the magnetic slab. As an illustrative example, we will use the solar corona conditions. Before doing this, we should specify the cooling $L(\rho,T)$ and heating $H(\rho,T)$ rates.

The  cooling  $L(\rho,T)$ in the coronal conditions is mainly due to optically thin radiation:
\begin{equation}
L(\rho,T)=\frac{\rho}{4m^2}\Lambda\!\left(T\right),
\label{Cooling}
\end{equation}
where $\Lambda\!\left(T\right)$ is a function of radiation losses depending on the plasma temperature. In this paper, the function $\Lambda\!\left(T\right)$ is calculated from the CHIANTI Version 10.0 database.

The heating  rate $H(\rho,T)$  is usually modeled by the power dependence on the thermodynamic parameters of the plasma \citep{Rosner1978,Dahlburg1988,Ibanez1993}:
\begin{equation}
H(\rho,T)=h \rho^aT^b,
\label{Heating}
\end{equation}
where $h$ is a constant calculated in order to balance cooling under steady state conditions $ (H(\rho_0,T_0)=L(\rho_0,T_0))$; $a$ and $b$ are the constants determined by a specific heating mechanism. 
{In general, a different form of the heating function will result in a different way for some initial perturbation to evolve \citep[see][for details]{2021SoPh..296...96Z}. Particularly, there are some forms of heating mechanisms implying the amplification of MA waves leading to formation of quasi-periodic perturbations \citep[see][]{Zavershinskii2019,2019A&A...628A.133K}. There are also some regimes of the thermal misbalance where not only MA waves but also  entropy/thermal modes are unstable \citep[see][]{Claes2019}.  However, analysis of the unstable regimes modes is beyond the scope of this research. It is a separate problem that will be addressed in our future studies.}

For our calculations, we will use the heating scenario seismologically proposed by \cite{Kolotkov_2020}  using the observations of the damped slow magnetoacoustic waves in long-lived coronal plasma structures. This mechanism is based on the assumption that the thermal stability (attenuation of thermal mode) $\tau_{V,P}>0$ and acoustic stability (attenuation of MA modes) $\left(\tau_{P}-\tau_{V}\right)/\tau_{P}\tau_{V} >0$ conditions are always satisfied in the coronal plasma. The power indices of heating function (\ref{Heating}) for this scenario are $(a = 0.5, b = - 3.5)$.

\subsection{Dispersion properties}

{The physical conditions of the coronal loops can vary widely in density contrast, thickness, magnetic field strength, temperature and etc. In this paper, we will limit our discussion to consideration of the three sets of  parameters corresponding to some different loop types. Thus, in what follows,  we will consider a magnetic slab with the parameters shown in Table \ref{parameters}}.

Let us introduce the characteristic scale which can be associated with the thermal misbalance. Using infinite magnetic field approximation, \cite{Zavershinskii2019} showed that  the effect of  dispersion caused by the thermal misbalance  is most pronounced when the wave period is about the characteristic times $\tau_{V,P}$ and  reaches its maximum near the frequency: 
 \begin{equation} 
	\omega_M = (\tau_{V} \tau_{P})^{-{\frac{1}{2}}}.
	\label{w_m}
\end{equation} 

In our calculations, we assume that heating/cooling processes act both inside and outside the slab. However, due to the density contrast, the cooling (\ref{Cooling})  and heating (\ref{Heating}) rates outside the slab are weaker. As we consider the heating/cooling rates as power law functions, we can find that the characteristic times $\tau_{V,P}$ (\ref{characteristic_times}) outside the slab are { 2, 5, and 10 of the value inside the slab for the chosen density contrast of ''cool'', ''warm'', and ''hot'' loops, respectively.} Thus, the impact of the thermal misbalance from the external plasma is suppressed. We choose the dimensionless wavenumber $k_{Z_M}=Re\left(k_{z}\right)x_0$ calculated by solving equation (\ref{Solution}) as a spatial scale associated with the thermal misbalance. For solution, we use  $\omega_{M_i}$, which is the frequency of the maximum dispersion effect in the internal plasma.

{Generally, the phase velocities of the fast and slow waves in the solar corona vary in the weakly comparable ranges. This is due to the fact that plasma beta in such medium is usually less than unity, which implies a strong difference between the Alfv\'en and the sonic speeds. For this reason, we will use two spatial scales on the phase velocity plots to illustrate of the fast and slow wave dispersion. }
\begin{figure*}
	\begin{minipage}{0.49\linewidth}
		{\includegraphics[width=1.02\linewidth]{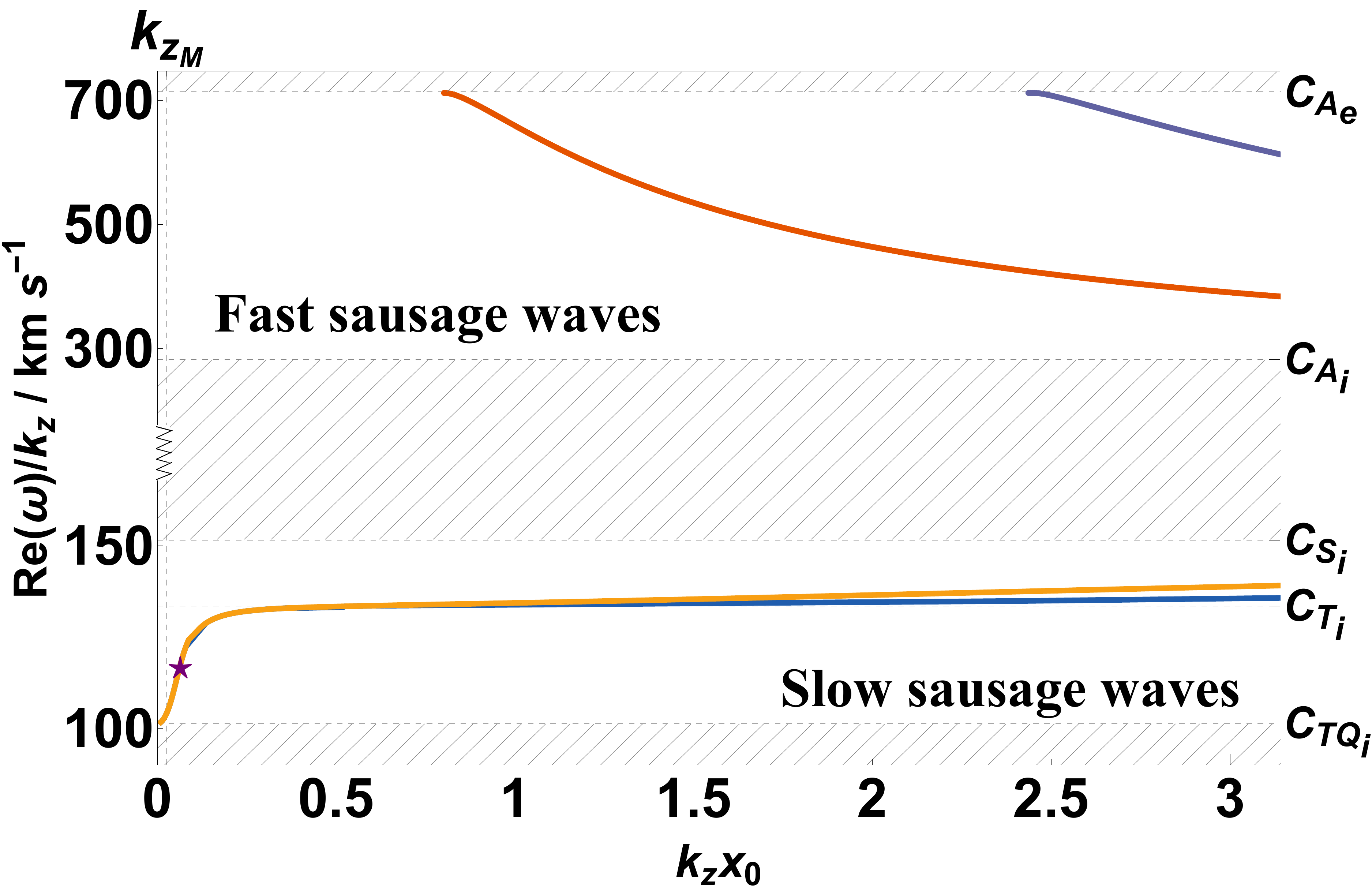}}
	\end{minipage}
	\hfill
	\begin{minipage}{0.49\linewidth}
		{\includegraphics[width=1.02\linewidth]{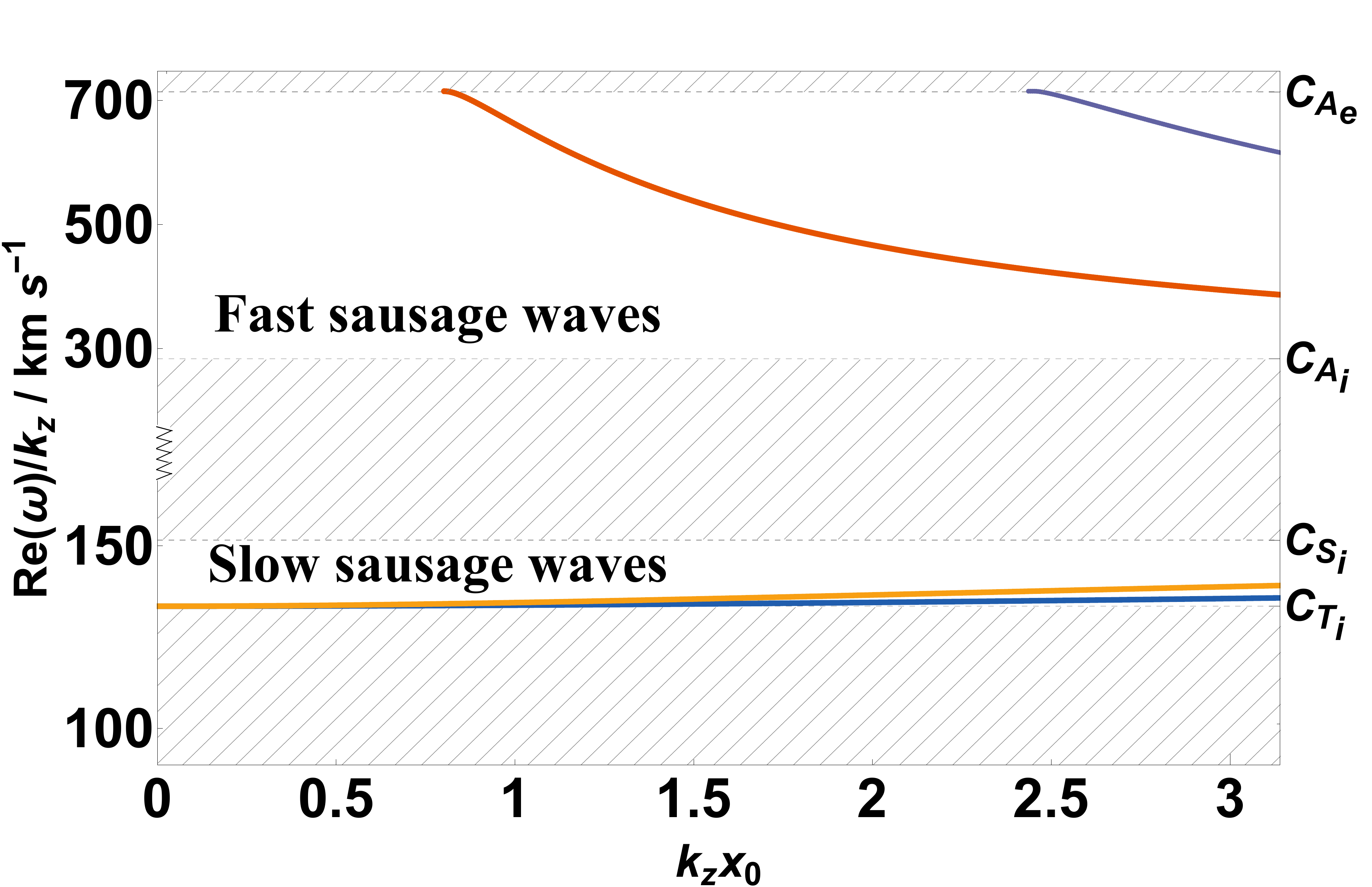}}
	\end{minipage}
	\begin{minipage}{0.49\linewidth}
		{\includegraphics[width=1.02\linewidth]{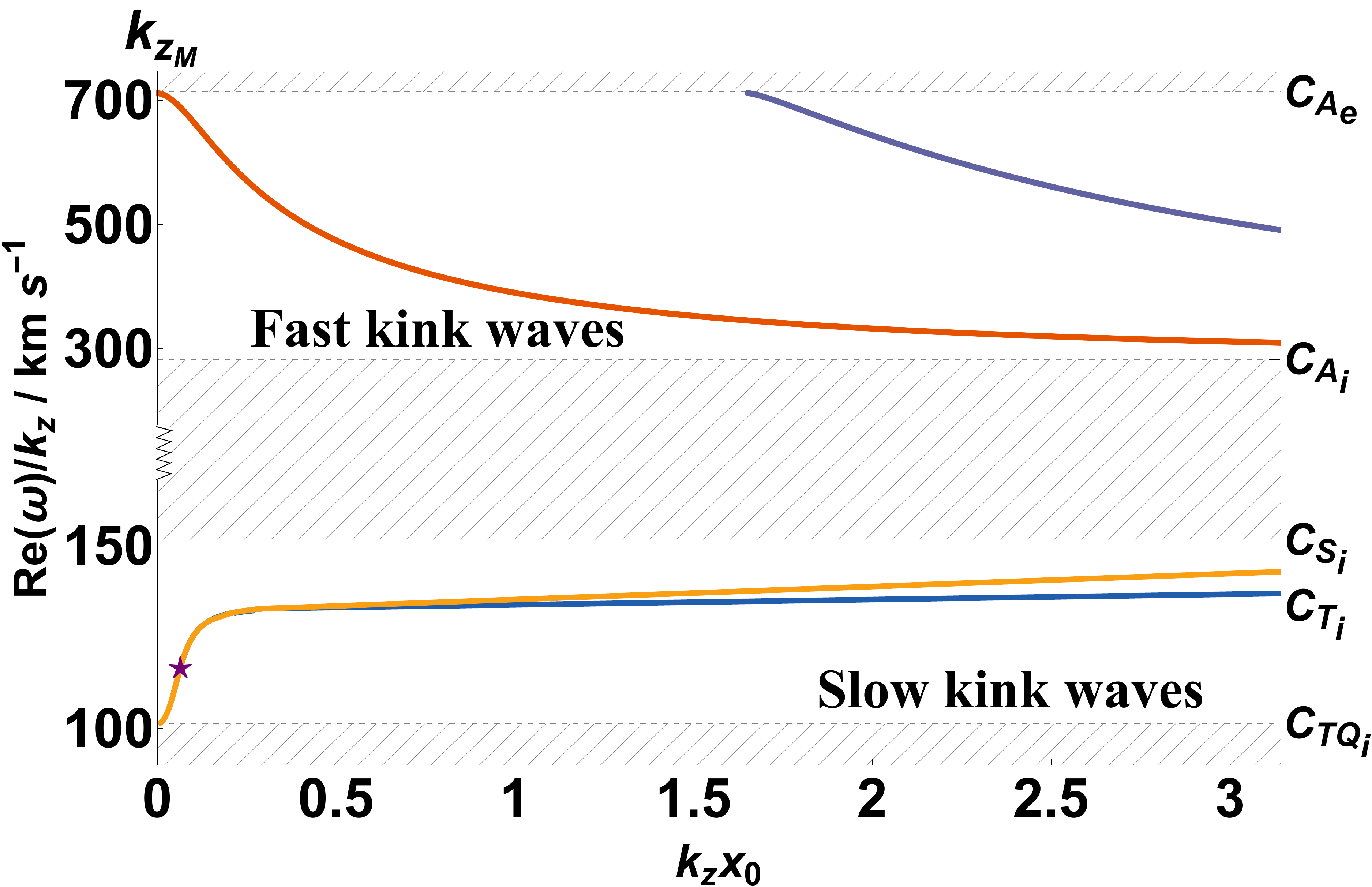}}
	\end{minipage}
	\hfill
	\begin{minipage}{0.49\linewidth}
		{\includegraphics[width=1.02\linewidth]{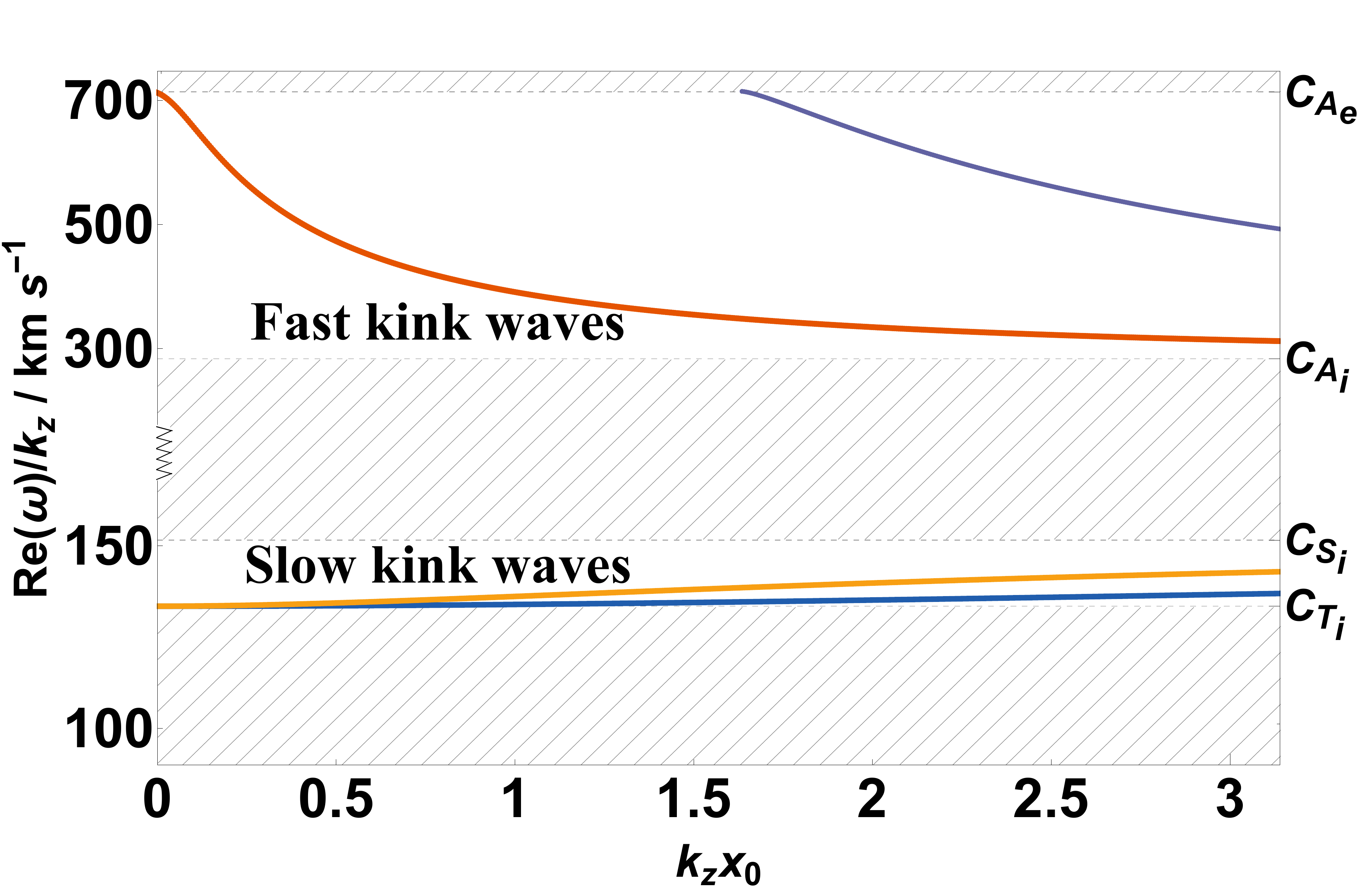}}
	\end{minipage}
	\caption{ Dependencies of the phase velocity $Re(\omega)/k$  on the  dimensionless wavenumber $kx_0$ {calculated for the ''warm'' coronal loop (see Table \ref{parameters}).} {We use different spatial scales for fast and slow MA waves. The range of velocities on vertical axis, where the scale is changing, is indicated by saw-teeth.}	The left column corresponds to the thermally active medium. The right panel shows the ideal plasma case. The top and bottom panels are for the sausage and kink modes, respectively. Different colors correspond to different modes. The star indicates the approximate position where the dispersion effect of slow waves is the most pronounced. Grey dashing indicates the range where no roots can be found. The slow modes in the thermally active plasma can be found between the sound speed $c_{S_i}$ and the modified tube speed $c_\mathrm{TQ_i}.$ The fast modes in the plasma with the thermal misbalance  range between $c_\mathrm{A_i}$ and $c_\mathrm{A_e}$.  }
	\label{Phase Velocity_Warm}
\end{figure*}
\begin{figure*}
	\begin{minipage}{0.49\linewidth}
		{\includegraphics[width=1.02\linewidth]{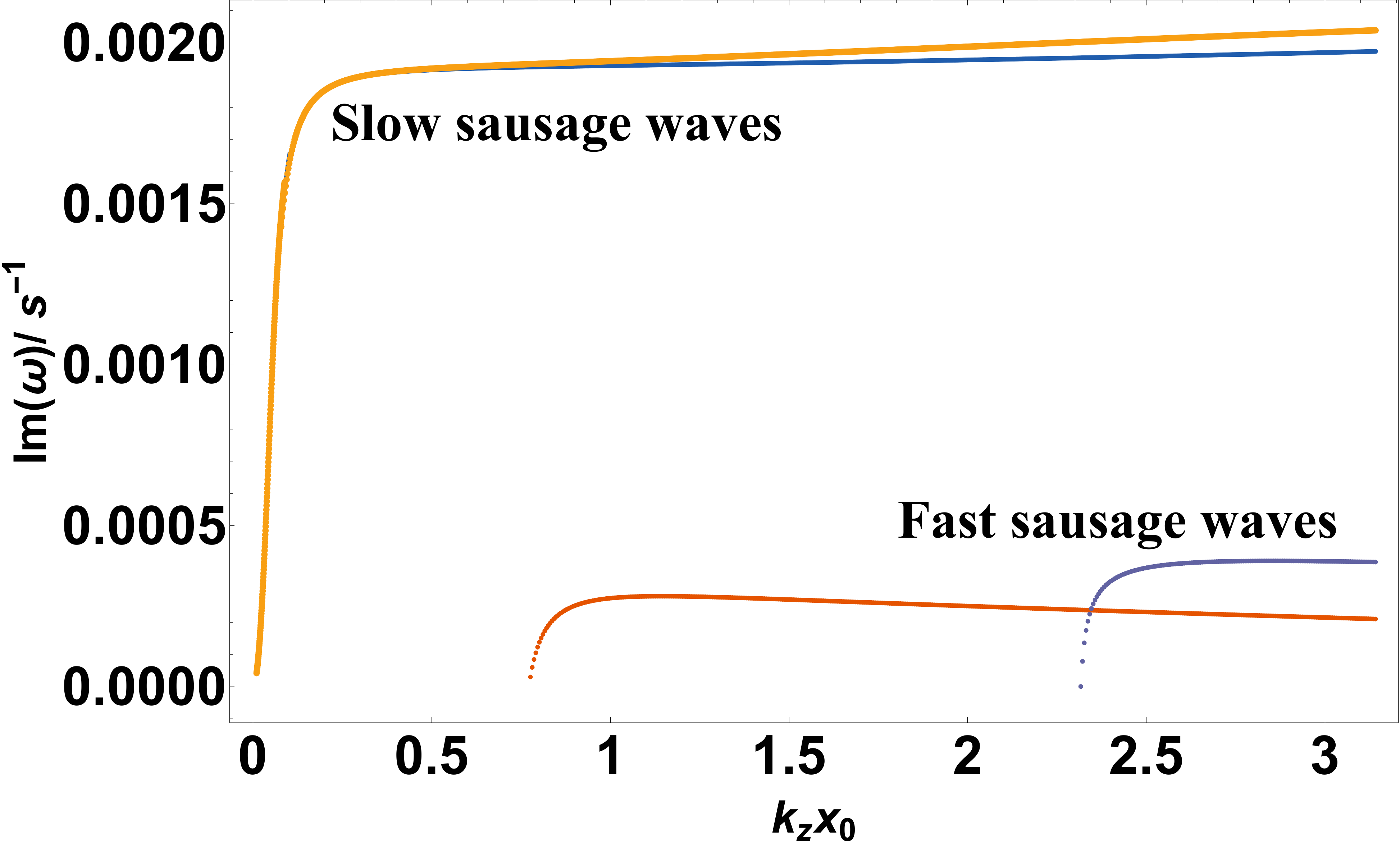}}
	\end{minipage}
	\hfill
	\begin{minipage}{0.49\linewidth}
		{\includegraphics[width=1.02\linewidth]{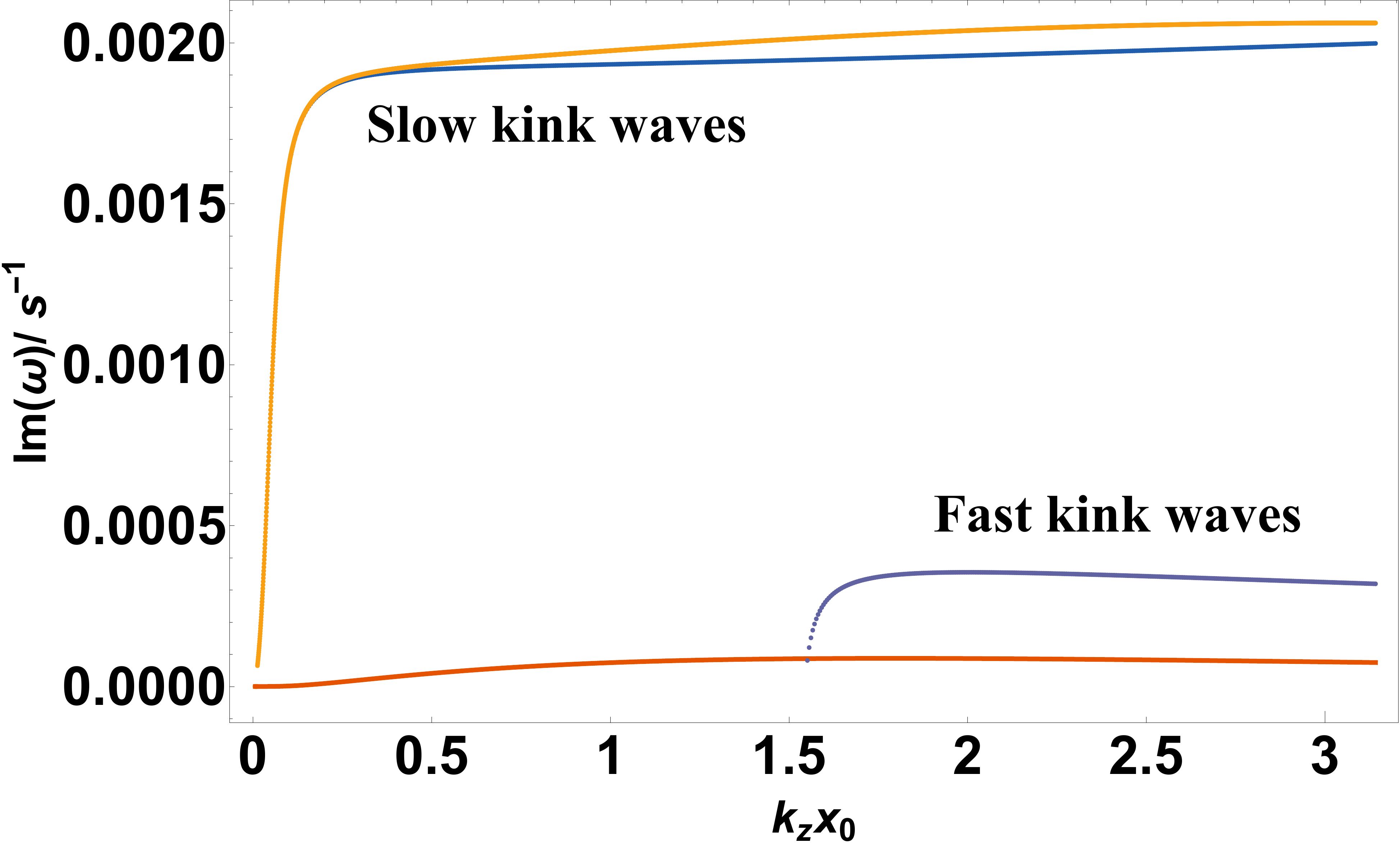}}
	\end{minipage}
	\caption{ Decrement $Im(\omega)$  as a function of  dimensionless wavenumber $kx_0$ {calculated for ''warm'' coronal loop (see Table \ref{parameters}).} The left column correspond to the sausage modes. The right panel is for the kink modes.  Different colors correspond to different modes.
	}
	\label{Fig_Im_W}
\end{figure*}
\subsubsection{ ''Warm'' loop}
\begin{figure*}
	\begin{minipage}{0.49\linewidth}
		{\includegraphics[width=1.02\linewidth]{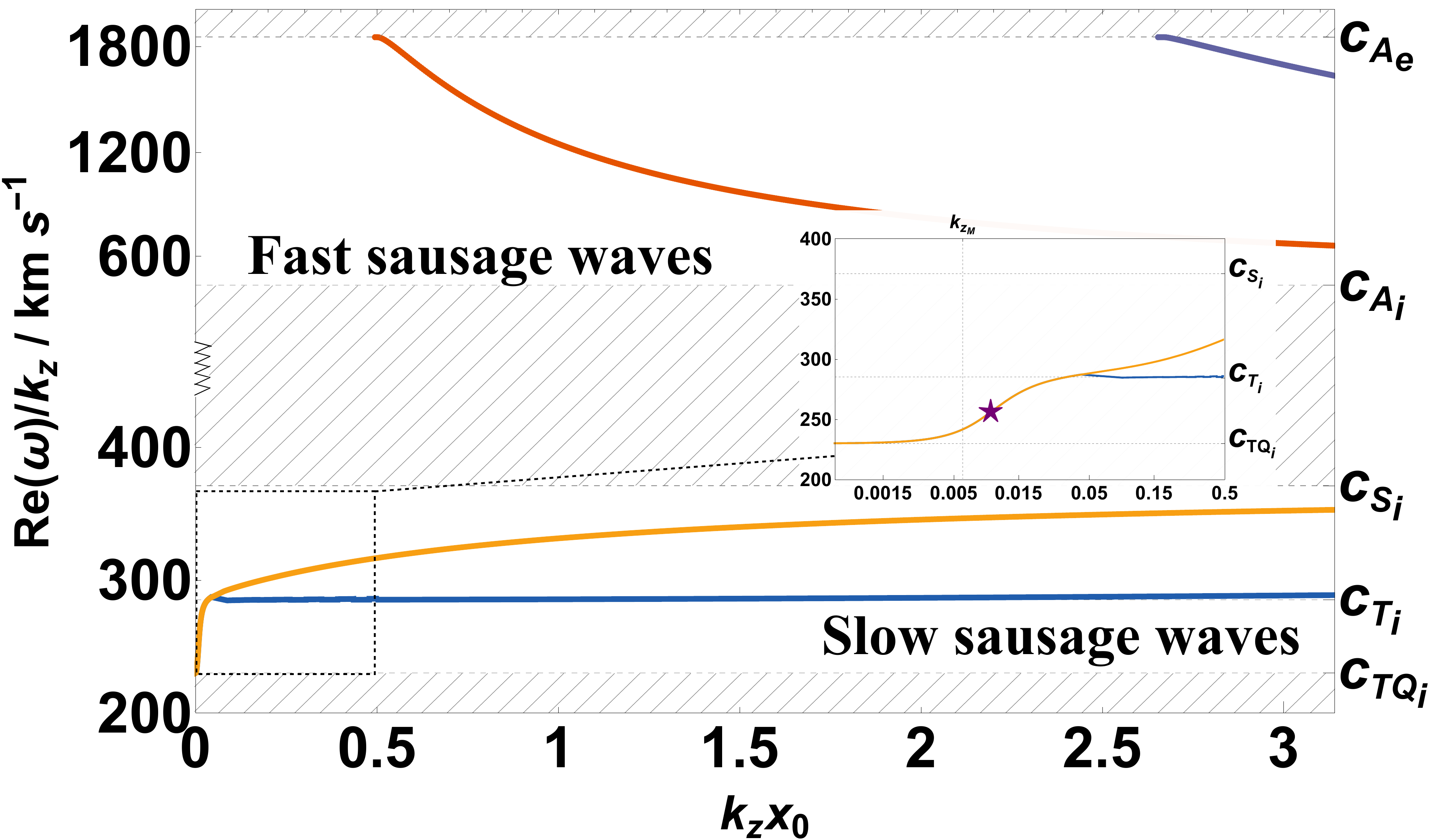}}
	\end{minipage}
	\hfill
	\begin{minipage}{0.49\linewidth}
		{\includegraphics[width=1.02\linewidth]{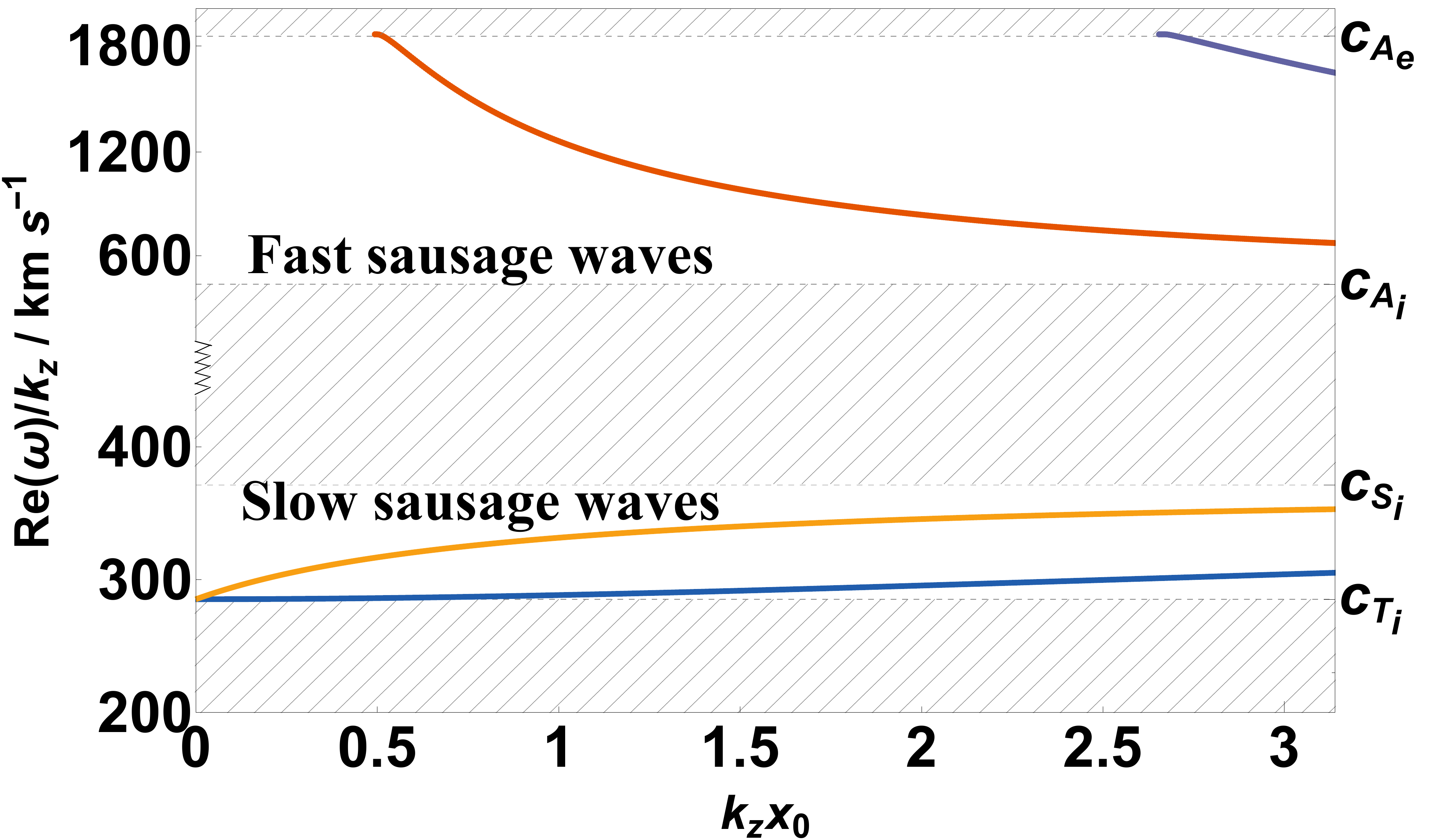}}
	\end{minipage}
	\begin{minipage}{0.49\linewidth}
		{\includegraphics[width=1.02\linewidth]{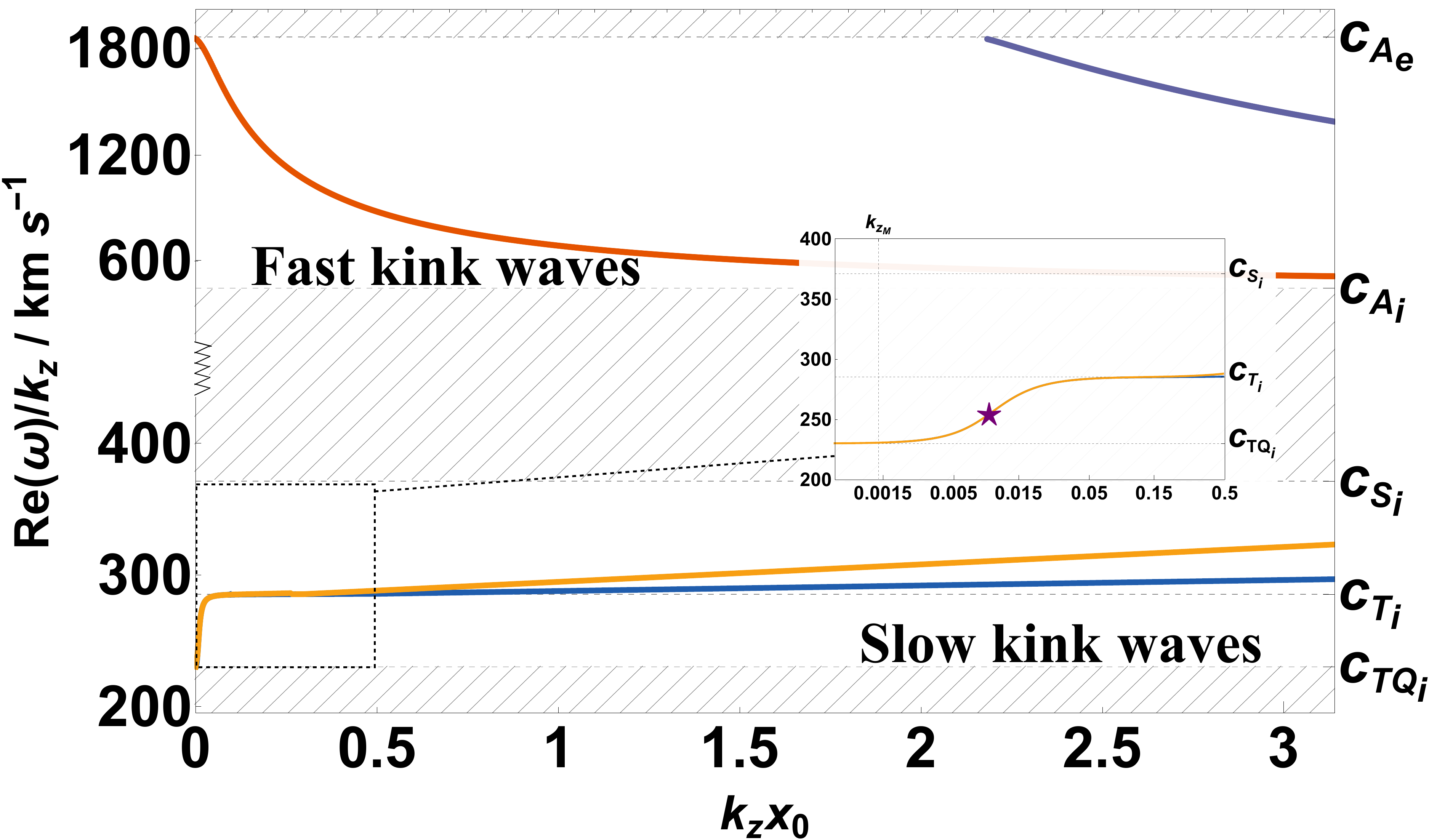}}
	\end{minipage}
	\hfill
	\begin{minipage}{0.49\linewidth}
		{\includegraphics[width=1.02\linewidth]{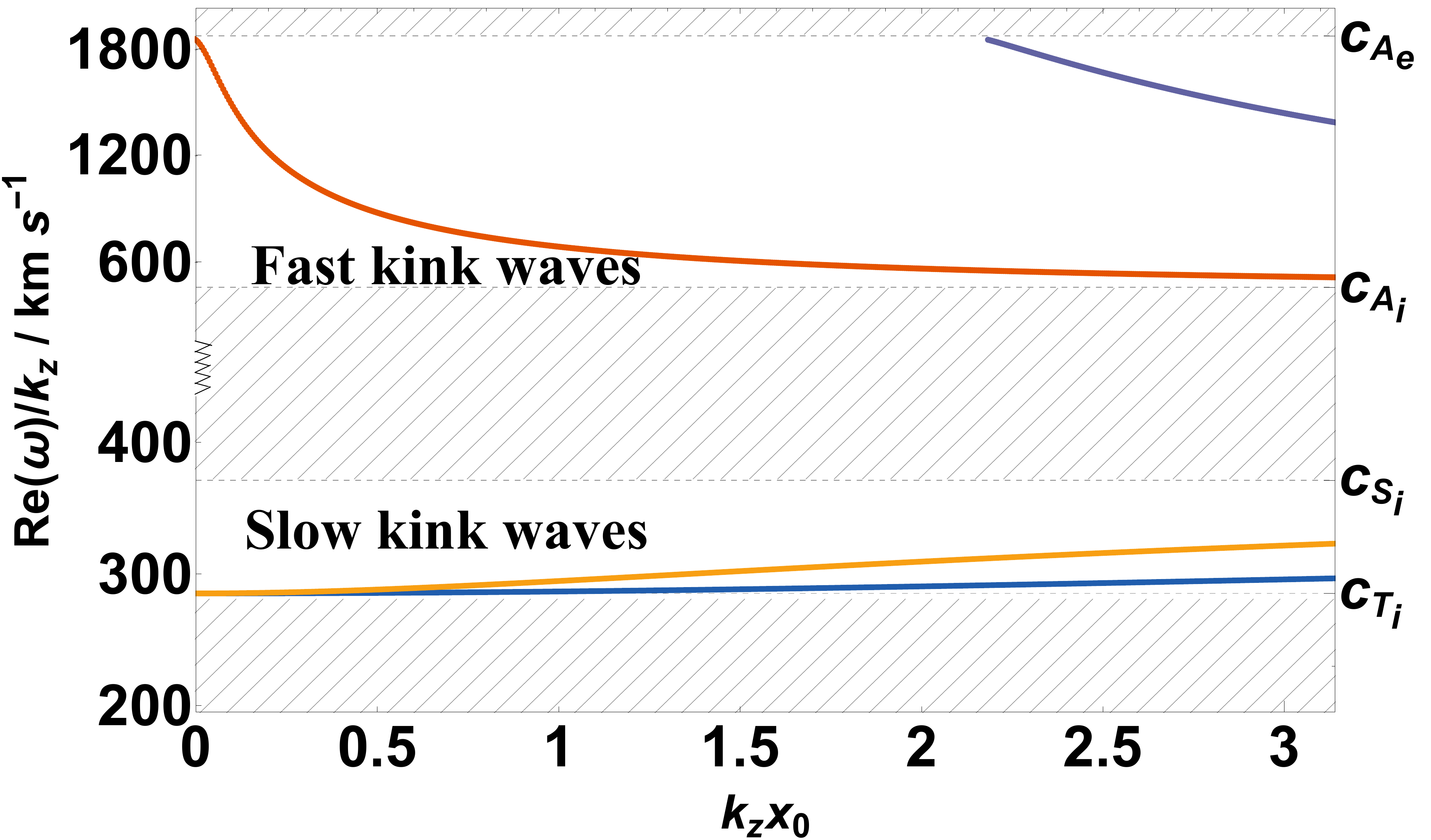}}
	\end{minipage}
	\caption{ Dependencies of the phase velocity $Re(\omega)/k$  on the  dimensionless wavenumber $kx_0$ {calculated for ''hot'' coronal loop (see Table \ref{parameters}).} {We use different spatial scales for the fast and slow MA waves. The range of velocities on vertical axis, where the scale is changing is indicated by saw-teeth.}	The left column corresponds to the thermally active medium. The right panel shows the ideal plasma case. The top and bottom panels are for the sausage and the kink modes, respectively. Different colors corresponds to different modes. The star indicates the approximate position where the dispersion effect of the slow waves is the most pronounced. Grey dashing indicates the range where no roots can be found. The slow modes in the thermally active plasma can be found between the sound speed $c_{S_i}$ and the modified tube speed $c_\mathrm{TQ_i}.$ The fast modes in the plasma with the thermal misbalance range between $c_\mathrm{A_i}$ and $c_\mathrm{A_e}$.  }
	\label{Phase Velocity_Hot}
\end{figure*}

\begin{figure*}
	\begin{minipage}{0.49\linewidth}
		{\includegraphics[width=1.02\linewidth]{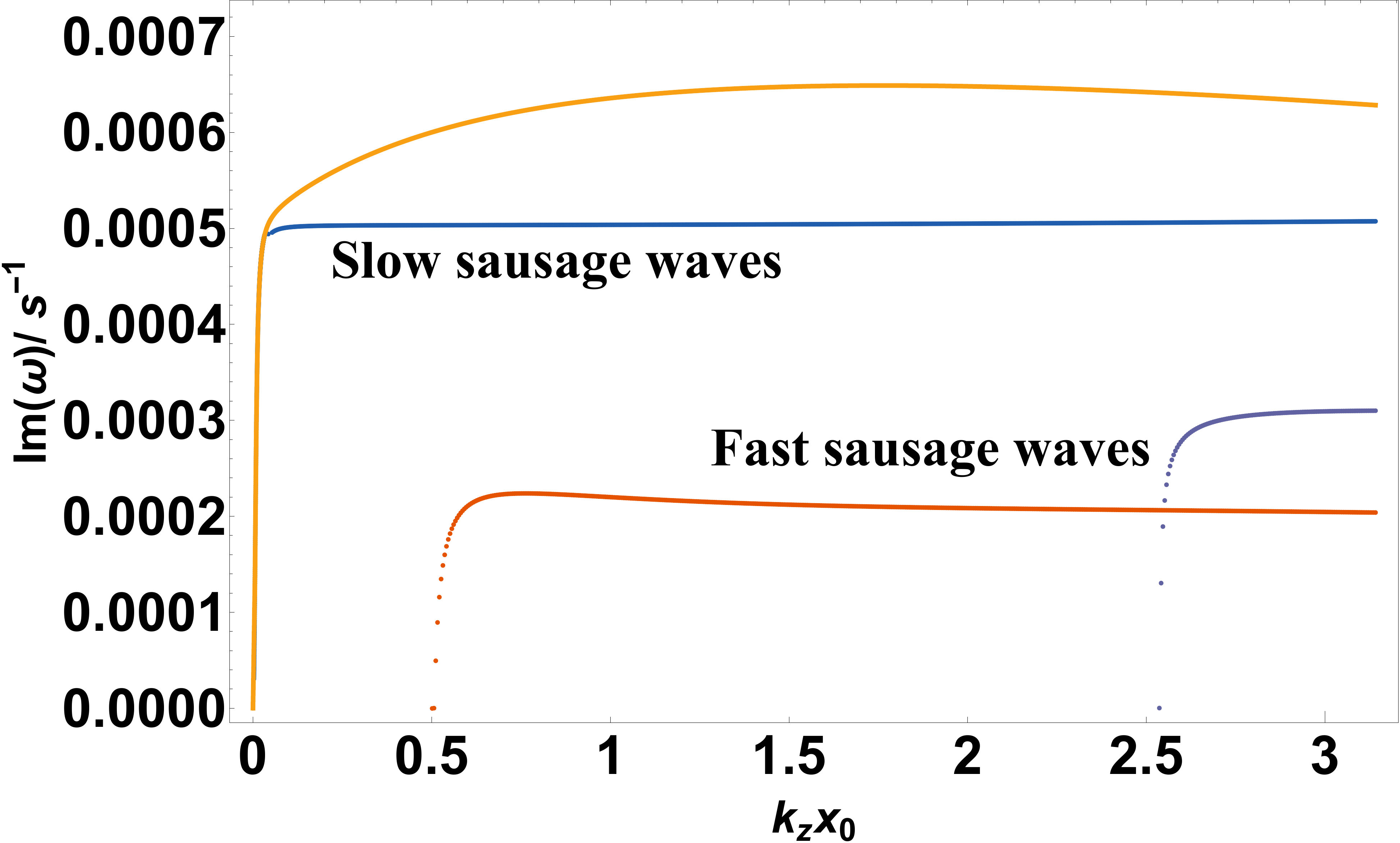}}
	\end{minipage}
	\hfill
	\begin{minipage}{0.49\linewidth}
		{\includegraphics[width=1.02\linewidth]{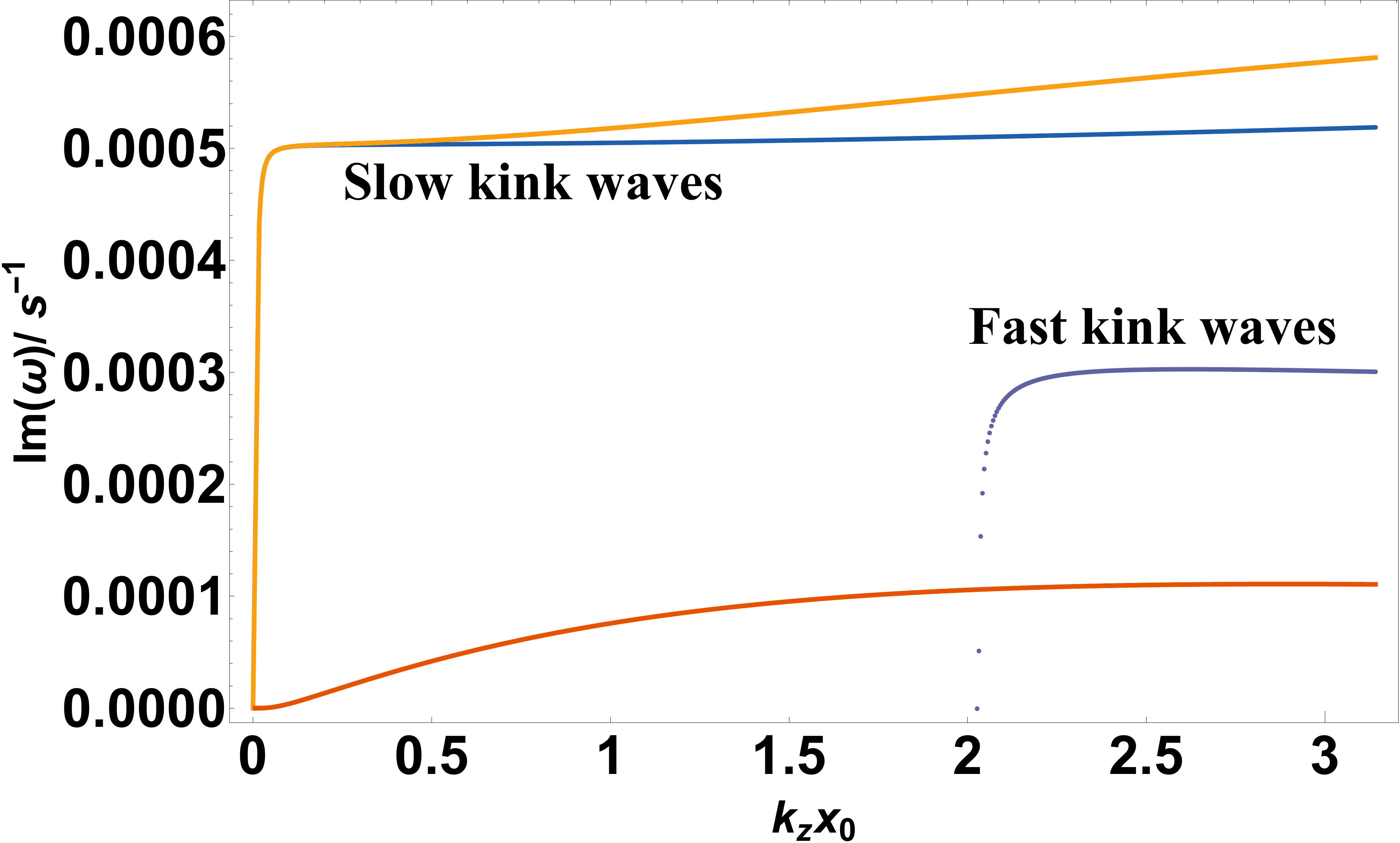}}
	\end{minipage}
	\caption{Decrement $Im(\omega)$  as a function of  dimensionless wavenumber $kx_0$ {calculated for ''hot'' coronal loop (see Table \ref{parameters}).} The left column correspond to the sausage modes. The right panel is for the kink modes.  Different colors correspond to different modes.
	}
	\label{Fig_Im_W_Hot}
\end{figure*}
The phase velocities of the fast and slow sausage/kink MA waves calculated for the sets of parameters corresponding to ''warm'' coronal loop are shown in Fig.~\ref{Phase Velocity_Warm}.  We consider both cases of the thermally active and ideal plasma in order to demonstrate their differences. According to our calculations, all found roots correspond to body waves.

Speaking of slow MA waves, one may notice that the thermal misbalance leads to considerable changes in the low-frequency (long-wavelength) range of the spectrum. It can be seen that the phase speed of the slow MA waves varies in the range between  $c_\mathrm{TQ_i} $ and $c_\mathrm{S_i} $. This statement is valid for both the kink and the sausage modes. For both symmetry types, the long-wavelength limit value of the phase speed $\lim\limits_{k_z x_0 \to 0} \mathrm{Re} (\omega)/ k$ is no longer  the usually assumed internal tube speed $c_\mathrm{T_i}$ but the modified tube speed $c_\mathrm{TQ_i} $ (\ref{characteristic_tube_speeds}). Moreover, the combination of two dispersion sources, namely, the geometry and   thermal misbalance, leads to a change in the period where the dispersion effect is most pronounced. Using calculation, it can be estimated as $\sim 700~s $ corresponding to $k_z x_0 \sim 0.075$, compared to $P_\mathrm{M}=2\pi/\omega_{M_i}\approx 860~s$ corresponding to $k_{Z_M} \approx 0.025 $ for the sausage waves ( indicated by star in Fig.~\ref{Phase Velocity_Warm}).

At the same time, the thermal misbalance seems to have no significant  affect on the fast MA waves. The phase speed dependencies on the wavenumber calculated for the ideal plasma and thermally active plasma are practically identical (see Fig.~\ref{Phase Velocity_Warm}).
In other words,  the slab geometry remains the main source of the phase speed dispersion for fast waves. Variations of the heating function satisfying the seismological constraints proposed in \citep{Kolotkov_2020} also do not lead to any considerable consequences either. As in the ideal plasma, the phase speed of the fast waves is varying between $c_\mathrm{A_i}$ and $c_\mathrm{A_e}$.

The thermal misbalance also  causes the frequency dependent dissipation of  waves (or amplification in the case of acoustic instability). The decrements of the fast and slow sausage/kink MA waves calculated for the chosen set of parameters are shown in Fig.~\ref{Fig_Im_W}. It can be seen that for the considered  regime of thermal misbalance all modes are decaying. 

It has been shown that the slow MA waves decay/grow faster than the fast MA waves in the uniform low-beta plasma \citep[see][for details]{2020PhRvE.101d3204Z}. The same is true for the magnetically structured plasma. However, there are some changes in the relationship between the increment/decrement and the wavenumber. In contrast to the case of the uniform plasma, the dependence of the fast wave increment/decrement becomes non-monotonic. It reaches some maximum value and then tends to zero in the high-frequency (short-wavelength) limit. The dependence of the slow-wave increment on the wavenumber remains monotonic.

\subsubsection{ ''Hot'' loop}

{For the plasma parameters corresponding to the ''hot'' loop (see the middle column in Table \ref{parameters}), the phase velocities of the MA waves take the form shown in Fig.~\ref{Phase Velocity_Hot}. As in the previous case, all the obtained fast and slow modes are body modes.
}
{Following the above line of reasoning, we will describe phase velocities first and start with slow waves. Due to a slight increase in the plasma beta, the variations of slow waves phase velocities caused by geometrical effects become more visible (compare the plots shown in the right columns in Fig. ~\ref{Phase Velocity_Warm} and ~\ref{Phase Velocity_Hot}).  The impact of geometry dispersion is more pronounced for the sausage waves  than for kink waves. }

{One may notice that for the considered plasma parameters, the combined effect of geometrical and thermal dispersion (indicated by  star in  Fig.~\ref{Phase Velocity_Hot} ) is most pronounced around $k_z x_0 \sim 0.0095$, which corresponds to the period $\sim 2600~s $. Such an increase in the dispersion timescale is caused by changes in the characteristic times $\tau_V, \tau_P$ (\ref{characteristic_times}). The misbalance timescales, in turn, are defined by the derivatives of heating (\ref{Heating}) and cooling (\ref{Cooling}) rates, and as a consequence, they are quite sensitive to the absolute value and slope of these functions. For the sake of brevity, we omit the details of changing the derivatives of functions. However, we want to emphasize that during the estimations one should take special attention to a strong dependence of the cooling rate on temperature   $\Lambda\!\left(T\right)$ \citep[see][]{2021ApJ...909...38D} and the choice of a heating model, as the variations in estimations could be dramatic. 
}

\begin{figure*}
	\begin{minipage}{0.49\linewidth}
		{\includegraphics[width=1.02\linewidth]{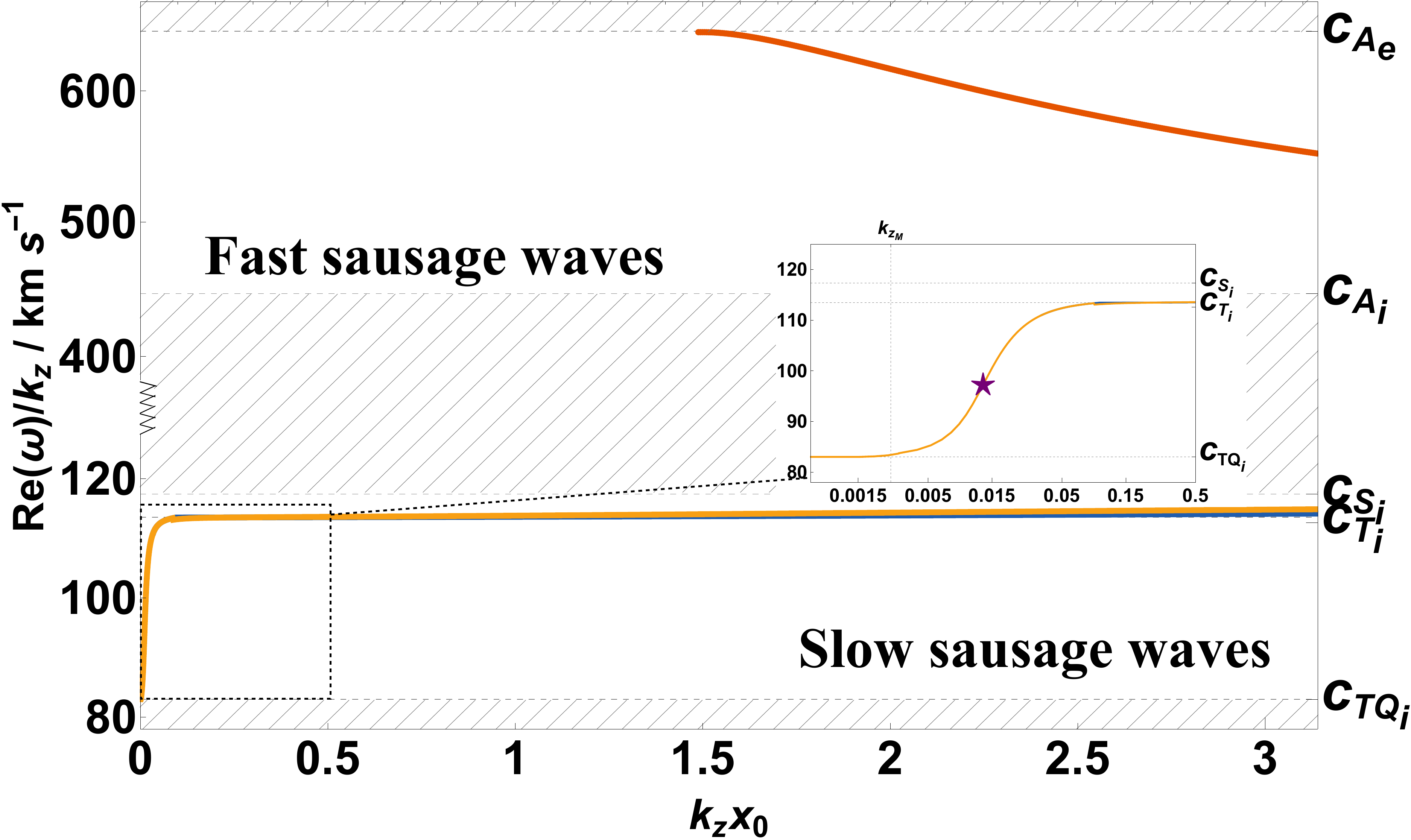}}
	\end{minipage}
	\hfill
	\begin{minipage}{0.49\linewidth}
		{\includegraphics[width=1.02\linewidth]{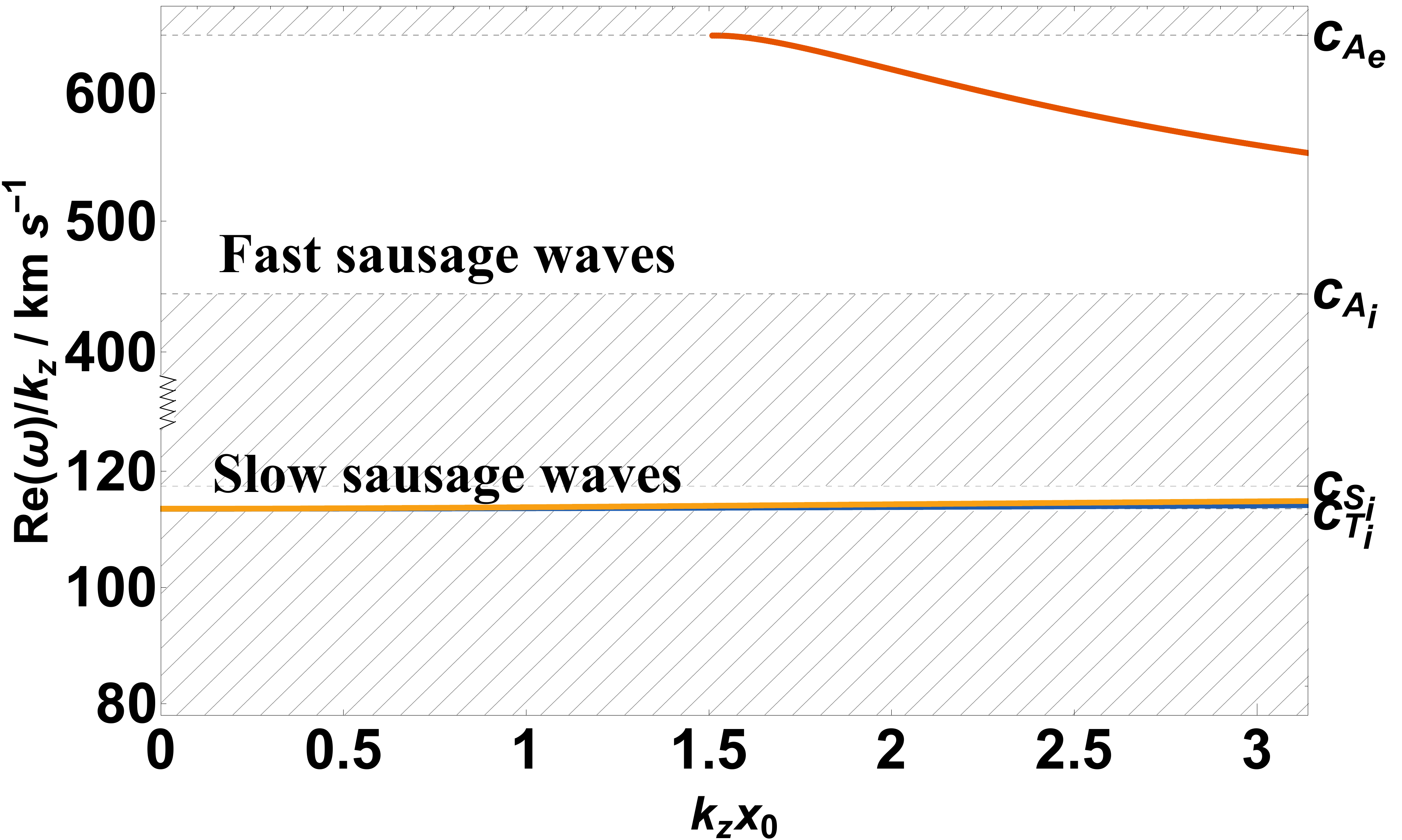}}
	\end{minipage}
	\begin{minipage}{0.49\linewidth}
		{\includegraphics[width=1.02\linewidth]{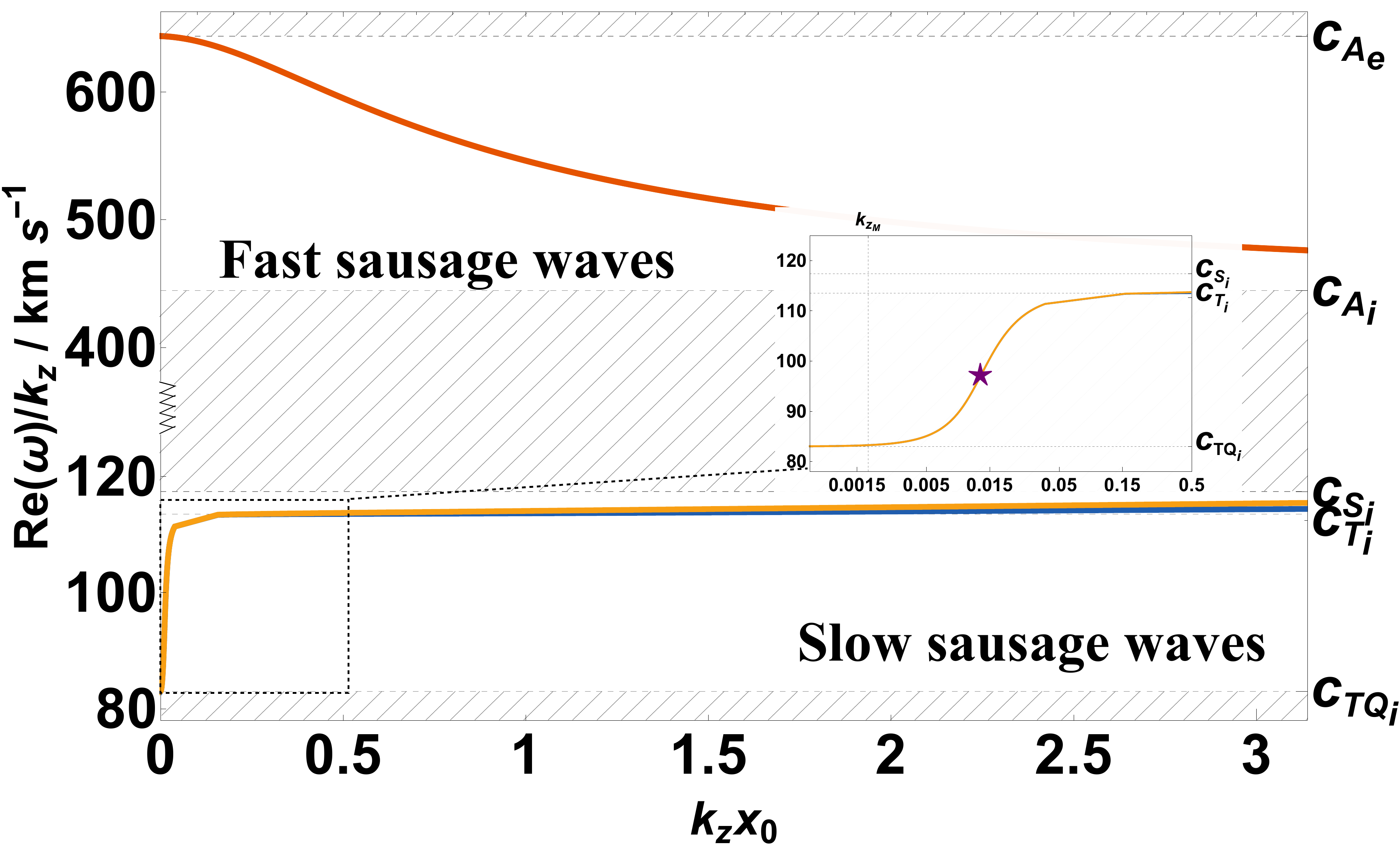}}
	\end{minipage}
	\hfill
	\begin{minipage}{0.49\linewidth}
		{\includegraphics[width=1.02\linewidth]{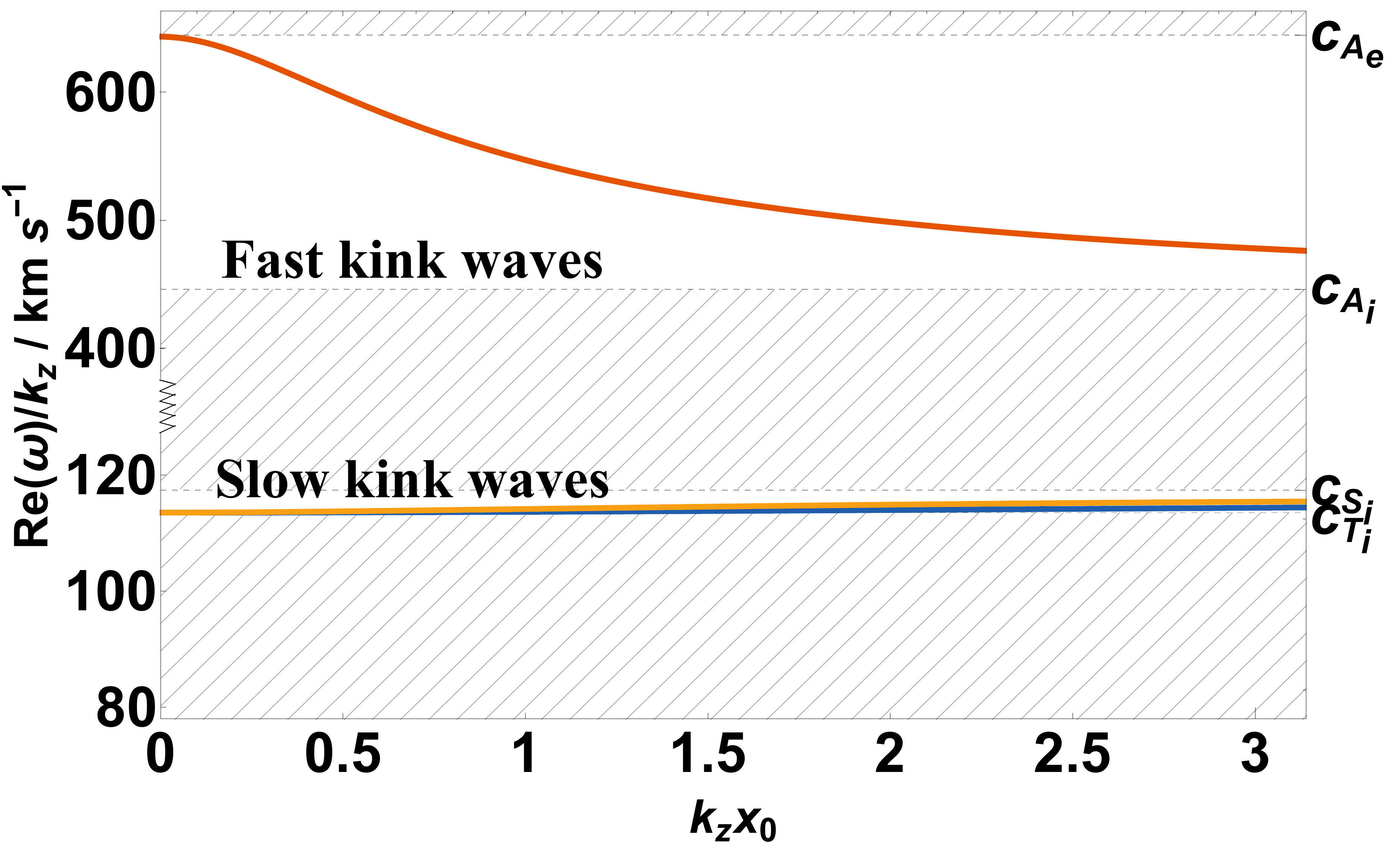}}
	\end{minipage}
	\caption{ Dependencies of the phase velocity $Re(\omega)/k$  on the  dimensionless wavenumber $kx_0$ {calculated for the ''cool'' coronal loop (see Table \ref{parameters}).} {We use different spatial scales for the fast and slow MA waves. The range of velocities on vertical axis, where the scale is changing, is indicated by saw-teeth.}	The left column corresponds to the thermally active medium. The right panel shows the ideal plasma case. The top and bottom panels are for the sausage and the kink modes, respectively. Different colors corresponds to different modes. The star indicates the approximate position where the dispersion effect of slow waves is the most pronounced. Grey dashing indicates the range where no roots can be found. The slow modes in the thermally active plasma can be found between the  sound speed $c_{S_i}$ and the modified tube speed $c_\mathrm{TQ_i}.$ The fast modes in the plasma with the thermal misbalance are in the range between $c_\mathrm{A_i}$ and $c_\mathrm{A_e}$.  }
	\label{Phase Velocity_Cool}
\end{figure*}

\begin{figure*}
	\begin{minipage}{0.49\linewidth}
		{\includegraphics[width=1.02\linewidth]{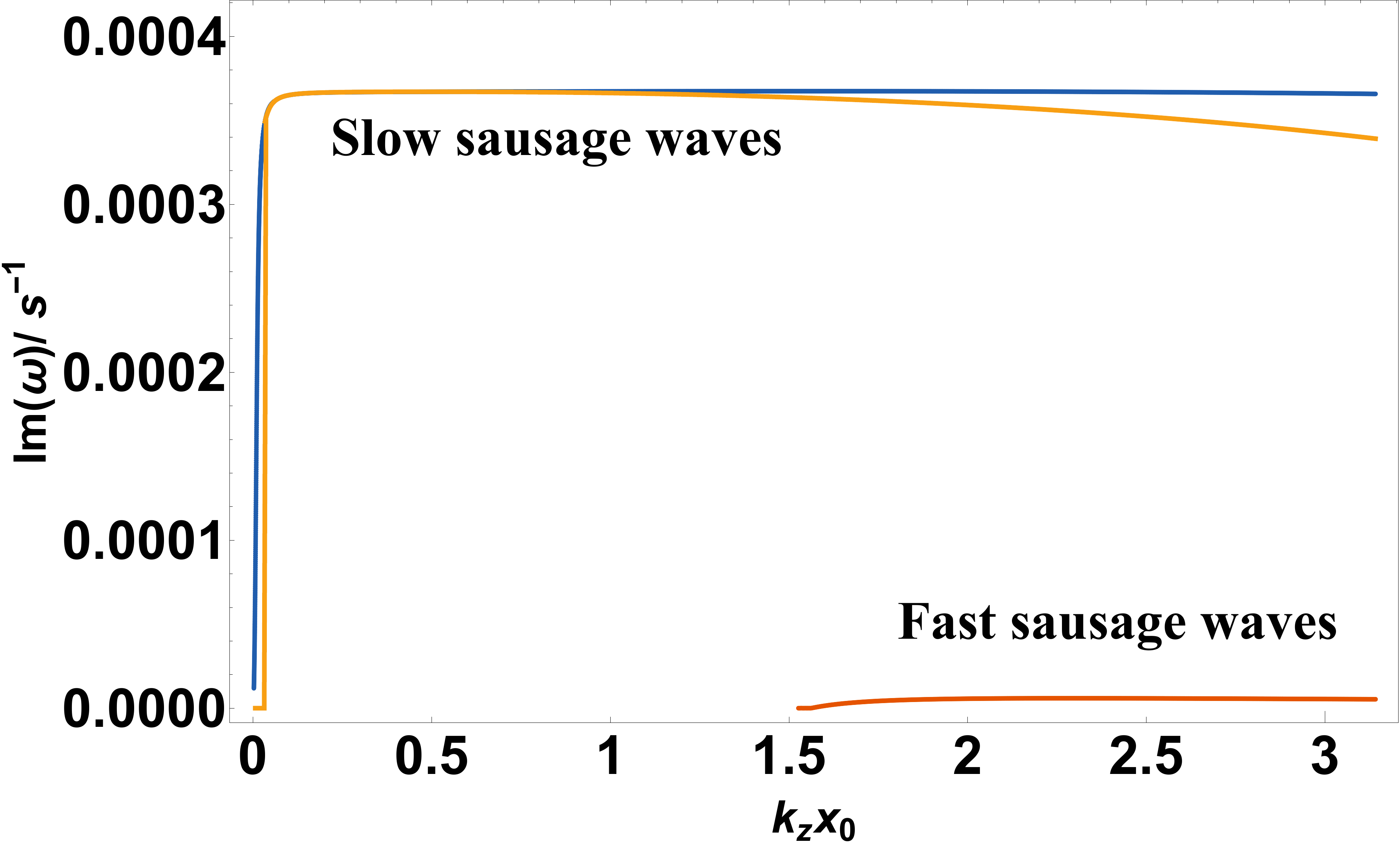}}
	\end{minipage}
	\hfill
	\begin{minipage}{0.49\linewidth}
		{\includegraphics[width=1.02\linewidth]{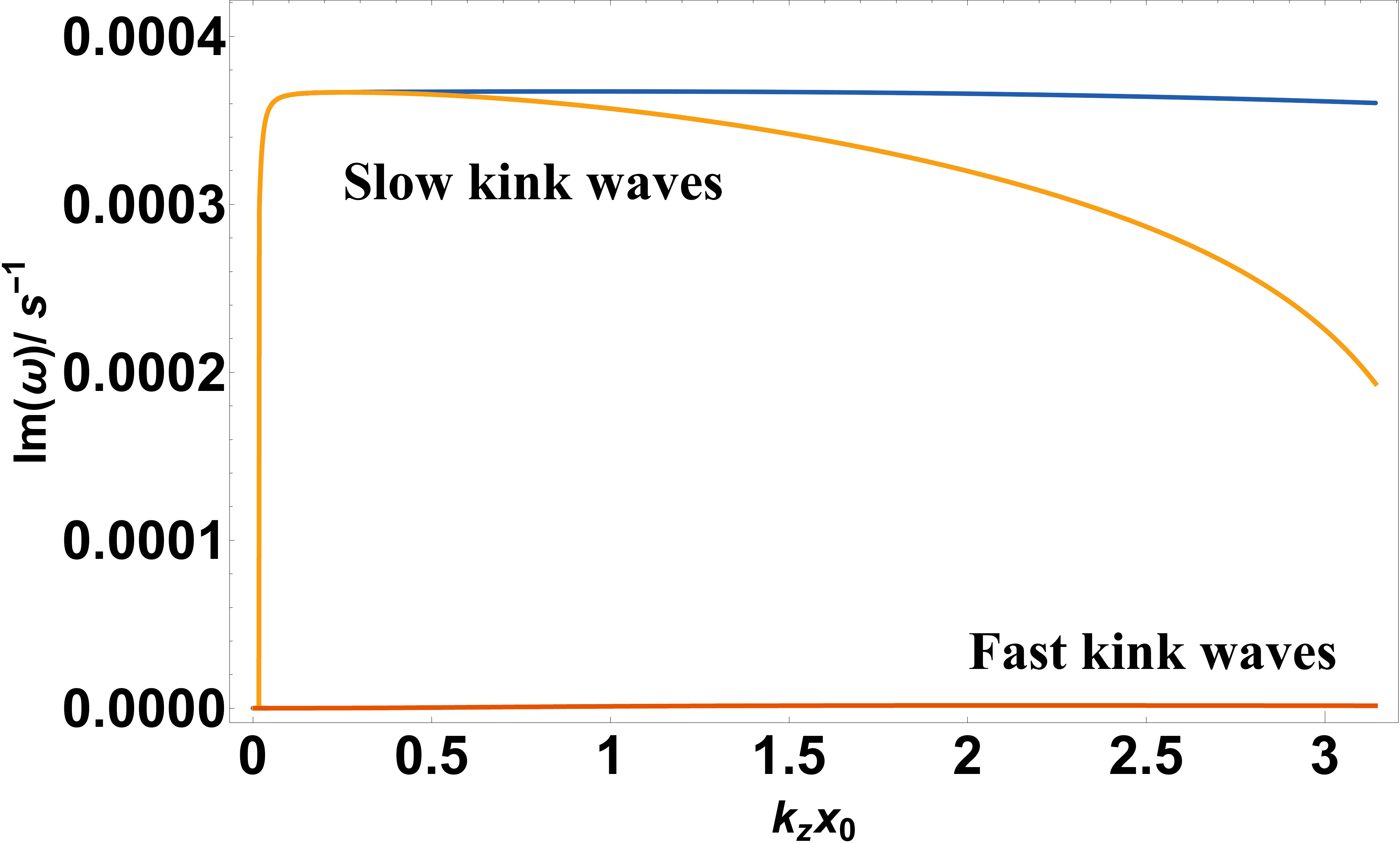}}
	\end{minipage}
	\caption{Decrement $Im(\omega)$  as a function of  dimensionless wavenumber $kx_0$ {calculated for the ''cool'' coronal loop (see Table \ref{parameters}).} The left column correspond to the sausage modes. The right panel is for the kink modes.  Different colors correspond to different modes.
	}
	\label{Fig_Im_W_Cool}
\end{figure*}

{It can be seen that the thermal misbalance still have no significant impact on the phase velocities of the fast MA waves.}

{For the considered parameters of ''hot'' coronal loop, the fast and slow waves are still decaying modes (see Fig.\ref{Fig_Im_W_Hot}). As a result of the above mentioned slight increase of the plasma beta, the difference between absolute values of the slow wave and fast wave decrements decreases. This is in agreement with the predictions made for the MA waves in the uniform plasma \citep[see][]{2020PhRvE.101d3204Z}. It should be noted that for the case of ''hot'' loop, not only the fast waves have a non-monotonic dependence of the decrement on the wavenumber. The decrement of the slow sausage wave shows a non-monotonic behaviour as well. However, as in the case of the ''warm'' loop, the decrement of the kink waves grows monotonically with the wavenumber.  }

\subsubsection{ ''Cool'' loop}

{Here we are dealing with the influence of the thermal misbalance and magnetic structuring on the MA waves using the  plasma parameters corresponding to the ''cool'' loop (see the right column in Table \ref{parameters}). The phase velocities calculated for these parameters are shown in Fig.~\ref{Phase Velocity_Cool}. }

{Let us, as before, start with the phase velocity of the slow waves. The used parameters correspond to the lower plasma beta than in the cases of ''warm'' and ''hot'' plasma. As a result, the difference between the ideal plasma sound  $c_\mathrm{S_i}$ and tube  $c_\mathrm{T_i}$ speeds decreases (compare the plots shown in the right columns in Figs. ~\ref{Phase Velocity_Warm}, ~\ref{Phase Velocity_Hot}  and ~\ref{Phase Velocity_Cool}). The thermal misbalance typically affects to a greater extent the speed of harmonics in the long-wavelength range of the spectrum. In the long-wavelength limit $( k_\mathrm{z} x_0 \to 0 )$, the phase speed exceeds the value $c_\mathrm{TQ_i} = 83\, km/s $ that is $30\,\% $ less than the usually assumed $c_\mathrm{T_i}$. The greatest increase in the slow wave phase speed with the wavelength (indicated by star  in  Fig.~\ref{Phase Velocity_Cool} )  is now seen around $k_z x_0 \sim 0.0125$, which corresponds to the period $\sim 5100~s $. In other words, for the considered heating mechanism, the ''warm'' coronal loops have the lowest period where the dispersion effect caused by thermal misbalance and magnetic structuring is most pronounced.}

{
In the case under consideration, the situation with fast MA waves is similar to those discussed before, i.e. the main source of the fast wave phase velocity dispersion remains the geometry of the wave-guide.} {As a result of the plasma beta decrease, the difference between the slow and fast wave decrement becomes greater (see Fig.\ref{Fig_Im_W_Cool}). This result also agrees well with the predictions made for the MA waves in the uniform plasma \citep[see][]{2020PhRvE.101d3204Z}. It can be seen that in the ''cool'' plasma conditions, the decrement behaviour of both the sausage and kink slow waves becomes non-monotonic.}

\section{Discussion and conclusion}
\label{s:Discussion}

Current studies into the thermal misbalance effect as applied to the solar atmosphere conditions are mainly proceeding from the assumptions related to the wave-guide geometry. The most commonly used approach involves applying the infinite magnetic field approximation \citep[see][etc]{Kolotkov_2020,2021SoPh..296..105P}. Encouraging results were obtained using the second or zero order thin flux tube approximation  \citep[see, e.g.,][]{Belov2021, 10.1088/1361-6587/ac36a5,Duckenfield2020effect}. However, these approaches are applicable for the description of the slow body sausage MA modes only. In this paper, we investigated the propagation of  the MA waves in a magnetically structured thermally active plasma without restrictions on the thickness of the waveguide or on the magnetic field strength. 

To describe the two-dimensional compressional perturbations in the plasma with thermal misbalance, we have obtained a differential equation (\ref{Disp}) that allows us to analyze the evolution and to define the dispersion properties of waves in the magnetic slab with some arbitrary profile of density and magnetic field strength across the slab. Furthermore, it can be used to analyze the resonant absorption effect in the non-adiabatic plasma, which is one of major problems that need to be addressed.

To investigate the dispersion properties of the MA waves, we have considered the case of a strong magnetic structuring (the step-function profiles of the density and magnetic field strength) in the slab geometry. The general approach used in this paper makes it possible to describe various types of wave modes. The dispersion relations for the fast/slow body/surface kink/sausage MA waves are written in the form of equations  (\ref{Solution}). This result widens the spectrum of modes that can be investigated in the context of the thermal misbalance problem (the problem of the coronal heating/cooling influence on the wave properties). We should remind that in the thermally active plasma, the modes are not purely body or purely surface modes but  ''rather body  than surface'' or ''rather surface than body'', respectively.

First, we solved  the relations  (\ref{Solution})  numerically using the parameters corresponding to coronal conditions. {In our calculations, we are using the plasma parameters corresponding to the three different loop types. } The found modes for the considered conditions are body modes implying ''rather body than surface'' modes. Next, we  separated the results obtained for the fast and slow waves.

As far as the slow waves are concerned, we have found out that, in contrast to the ideal plasma case, the thermal misbalance can change the slow-wave phase speed dependence. In particular, their phase speed tends to the  modified tube speed $c_\mathrm{TQ_i} $ (\ref{characteristic_tube_speeds}), which was previously defined in \citep{Belov2021} using the thin flux tube approximation. The modified tube speed  $c_\mathrm{TQ_i} $ is a consequence of both thermal-misbalance and finite-slab-width effects. In this paper, we have established that this result is  applicable for both the sausage and  the kink slow modes. {The tendency is true for all the plasma parameters we considered. However,  the period when the dispersion effects caused by the combination of magnetic structuring and thermal misbalance are  most pronounced, is shorter in the ''warm'' loops ($\approx 860~s$), than in the ''cool'' ($\approx 5100~s$) and ''hot'' ($\approx 2600~s$) loops.}  This tendency means that propagation of the slow-wave harmonics close to the fundamental mode is primarily influenced by the heating and cooling processes. As a result, the effective adiabatic index for these harmonic is different from the standard value. This result is of interest in the context of the observed temperature dependence of coronal adiabatic index \citep{2018ApJ...868..149K,2011ApJ...727L..32V} and its seismological definition using the  propagating \citep{2021SoPh..296..105P}  and standing  slow waves. Additionally, this phase dependency can be crucial for the probing magnetic field strength by propagating slow waves \citep{Jess2016}.

The heating mechanism considered in this paper implies  damping all compressional modes. We have established  that the damping rate of not only the sausage, but also of the kink modes depends on the wavenumber (frequency). { Moreover, behaviour of the wave decrement varies with the loop type. In the case of the ''warm'' loop,} the dependencies of increment/decrement on the wavenumber for both types of mode symmetry  is monotonic. The decrement of kink mode is slightly greater than the decrement of the sausage mode in the high-wavenumber range. As for the case of the infinite magnetic field approximation \citep{Zavershinskii2019}, the maximum decrement is reached in the range of high wavenumbers (frequencies). The minimum decay time is about  {8 minutes}, which coincides with the observed periods of waves and oscillations in the corona. Thus, the thermal misbalance can have an impact on the observed dissipation of slow waves propagating in the solar corona \citep{Marsh2011, Prasad2014}.

{In the case of the ''hot'' loop, the decrement of the slow kink waves grows with respect to the wavenumber. The minimum decay time is about  {30 minutes}. However, the sausage wave decrement shows a non-monotonic behaviour: it reaches the minimum decay time of $\sim 25$  minutes and then decreases (see the left column in Fig.~\ref{Fig_Im_W_Cool}). In the ''cool'' loop plasma, both the sausage and the kink slow waves demonstrates a non-monotonic behaviour of decrement. For both cases,  the minimum decay time is about  {46 minutes}.}

The results for the fast MA waves are quite different. Our analysis has revealed that the thermal misbalance caused by the temperature and density dependent heating/cooling processes has no significant effect on the phase speed of the kink and sausage fast MA waves. Thus, even in the non-adiabatic plasma, the main source of the fast-wave phase speed dispersion is the geometry of the slab.

The thermal activity of the plasma provides a possibility for the fast kink/sausage MA waves to be damped/amplified depending on the acting misbalance regime.  As it was obtained for the waves in the uniform plasma \citep[see][for details]{2020PhRvE.101d3204Z}, the fast waves have a lower decrement than the slow waves in the low-beta plasma. This statement is valid for the magnetically structured plasma as well. {Furthermore, our calculations have shown that the difference between the slow and fast wave decrement grows with the decrease in plasma beta, which is also in agreement with the results obtained for the uniform plasma.}
In contrast to the case of the uniform plasma, the dependence of the decrement on the wavenumber becomes non-monotonic: it reaches some maximum value and then tends to zero in the high-wavenumber range of spectrum. For the considered heating mechanism, the damping is rather weak and cannot explain the observed damping of the kink modes \citep[see, e.g.,][]{Pascoe2016}.
{The minimum decay time can be estimated as  {46 minutes}, and {55 minutes}, for the ''warm'', and ''hot''  loop, respectively. The influence of the thermal misbalance damping on the waves in the ''cool'' loop is negligible. } However, the obtained dependence of the fast wave decrement/increment is of great interest for the problem of fast kink oscillation excitation \citep[see][for details]{2021SSRv..217...73N}. The regime of acoustic instability will lead to the growth of perturbation harmonics with the fastest growth rate for the harmonics near the increment/decrement maximum. We will analyze this problem in our future study taking into account a possible dependence of the heating rate on the magnetic field strength and additional dissipation mechanisms.    
 
In conclusion, we want to emphasize that the constructed theory extends out knowledge of the properties and evolution of magnetoacoustic modes in the solar corona. This further understanding provides possibility to use these waves not only as a tool for the estimating of plasma parameters, but also as a tool for the estimating the non-adiabatic processes (e.g., for phenomenological determination of unknown coronal-heating mechanisms).


\section*{Acknowledgements}

The study was supported in part by the Ministry of Education and Science of Russia under State assignment to educational and research institutions under Project No. FSSS-2020-0014 and No. 0023-2019-0003, and by Russian Foundation for Basic Research, Project No. 20-32-90018. CHIANTI is a collaborative project involving George Mason University, the University of Michigan (USA), University of Cambridge (UK), and NASA Goddard Space Flight Center (USA).

\section*{Data Availability}

The data underlying this article will be shared on reasonable request
to the corresponding author.

\bibliographystyle{mnras}
\bibliography{refs} %

\begin{thebibliography}{}
\makeatletter
\relax
\def\mn@urlcharsother{\let\do\@makeother \do\$\do\&\do\#\do\^\do\_\do\%\do\~}
\def\mn@doi{\begingroup\mn@urlcharsother \@ifnextchar [ {\mn@doi@}
  {\mn@doi@[]}}
\def\mn@doi@[#1]#2{\def\@tempa{#1}\ifx\@tempa\@empty \href
  {http://dx.doi.org/#2} {doi:#2}\else \href {http://dx.doi.org/#2} {#1}\fi
  \endgroup}
\def\mn@eprint#1#2{\mn@eprint@#1:#2::\@nil}
\def\mn@eprint@arXiv#1{\href {http://arxiv.org/abs/#1} {{\tt arXiv:#1}}}
\def\mn@eprint@dblp#1{\href {http://dblp.uni-trier.de/rec/bibtex/#1.xml}
  {dblp:#1}}
\def\mn@eprint@#1:#2:#3:#4\@nil{\def\@tempa {#1}\def\@tempb {#2}\def\@tempc
  {#3}\ifx \@tempc \@empty \let \@tempc \@tempb \let \@tempb \@tempa \fi \ifx
  \@tempb \@empty \def\@tempb {arXiv}\fi \@ifundefined
  {mn@eprint@\@tempb}{\@tempb:\@tempc}{\expandafter \expandafter \csname
  mn@eprint@\@tempb\endcsname \expandafter{\@tempc}}}

\bibitem[\protect\citeauthoryear{{Banerjee} et~al.,}{{Banerjee}
  et~al.}{2021}]{2021SSRv..217...76B}
{Banerjee} D.,  et~al., 2021, \mn@doi [\ssr] {10.1007/s11214-021-00849-0},
  \href {https://ui.adsabs.harvard.edu/abs/2021SSRv..217...76B} {217, 76}

\bibitem[\protect\citeauthoryear{Belov, Molevich  \& Zavershinskii}{Belov
  et~al.}{2021}]{Belov2021}
Belov S.~A.,  Molevich N.~E.,   Zavershinskii D.~I.,  2021, \mn@doi [Solar
  Physics] {10.1007/s11207-021-01868-4}, 296

\bibitem[\protect\citeauthoryear{{Carbonell}, {Terradas}, {Oliver}  \&
  {Ballester}}{{Carbonell} et~al.}{2006}]{2006A&A...460..573C}
{Carbonell} M.,  {Terradas} J.,  {Oliver} R.,   {Ballester} J.~L.,  2006,
  \mn@doi [\aap] {10.1051/0004-6361:20065528}, \href
  {https://ui.adsabs.harvard.edu/abs/2006A&A...460..573C} {460, 573}

\bibitem[\protect\citeauthoryear{{Chin}, {Verwichte}, {Rowlands}  \&
  {Nakariakov}}{{Chin} et~al.}{2010}]{2010PhPl...17c2107C}
{Chin} R.,  {Verwichte} E.,  {Rowlands} G.,   {Nakariakov} V.~M.,  2010,
  \mn@doi [Physics of Plasmas] {10.1063/1.3314721}, \href
  {https://ui.adsabs.harvard.edu/abs/2010PhPl...17c2107C} {17, 032107}

\bibitem[\protect\citeauthoryear{{Cho} et~al.,}{{Cho}
  et~al.}{2017}]{2017ApJ...837L..11C}
{Cho} I.~H.,  et~al., 2017, \mn@doi [\apjl] {10.3847/2041-8213/aa611b}, \href
  {https://ui.adsabs.harvard.edu/abs/2017ApJ...837L..11C} {837, L11}

\bibitem[\protect\citeauthoryear{{Claes} \& {Keppens}}{{Claes} \&
  {Keppens}}{2019}]{Claes2019}
{Claes} N.,  {Keppens} R.,  2019, \mn@doi [\aap] {10.1051/0004-6361/201834699},
  \href {https://ui.adsabs.harvard.edu/abs/2019A&A...624A..96C} {624, A96}

\bibitem[\protect\citeauthoryear{{Dahlburg} \& {Mariska}}{{Dahlburg} \&
  {Mariska}}{1988}]{Dahlburg1988}
{Dahlburg} R.~B.,  {Mariska} J.~T.,  1988, \mn@doi [\solphys]
  {10.1007/BF00148571}, \href
  {https://ui.adsabs.harvard.edu/abs/1988SoPh..117...51D} {117, 51}

\bibitem[\protect\citeauthoryear{{Del Zanna}, {Dere}, {Young}  \& {Landi}}{{Del
  Zanna} et~al.}{2021}]{2021ApJ...909...38D}
{Del Zanna} G.,  {Dere} K.~P.,  {Young} P.~R.,   {Landi} E.,  2021, \mn@doi
  [\apj] {10.3847/1538-4357/abd8ce}, \href
  {https://ui.adsabs.harvard.edu/abs/2021ApJ...909...38D} {909, 38}

\bibitem[\protect\citeauthoryear{{Duckenfield}, {Kolotkov}  \&
  {Nakariakov}}{{Duckenfield} et~al.}{2021}]{Duckenfield2020effect}
{Duckenfield} T.~J.,  {Kolotkov} D.~Y.,   {Nakariakov} V.~M.,  2021, \mn@doi
  [\aap] {10.1051/0004-6361/202039791}, \href
  {https://ui.adsabs.harvard.edu/abs/2021A&A...646A.155D} {646, A155}

\bibitem[\protect\citeauthoryear{Edwin \& Roberts}{Edwin \&
  Roberts}{1982}]{Edwin1982}
Edwin P.,  Roberts B.,  1982, \mn@doi [Solar Physics] {10.1007/bf00170986}, 76

\bibitem[\protect\citeauthoryear{{Edwin} \& {Roberts}}{{Edwin} \&
  {Roberts}}{1983}]{1983SoPh...88..179E}
{Edwin} P.~M.,  {Roberts} B.,  1983, \mn@doi [\solphys] {10.1007/BF00196186},
  \href {https://ui.adsabs.harvard.edu/abs/1983SoPh...88..179E} {88, 179}

\bibitem[\protect\citeauthoryear{{Ibanez S.} \& {Escalona T.}}{{Ibanez S.} \&
  {Escalona T.}}{1993}]{Ibanez1993}
{Ibanez S.} M.~H.,  {Escalona T.} O.~B.,  1993, \mn@doi [\apj]
  {10.1086/173167}, \href
  {https://ui.adsabs.harvard.edu/abs/1993ApJ...415..335I} {415, 335}

\bibitem[\protect\citeauthoryear{{Jess} et~al.,}{{Jess}
  et~al.}{2016}]{Jess2016}
{Jess} D.~B.,  et~al., 2016, \mn@doi [Nature Phys] {10.1038/nphys3544}, \href
  {https://ui.adsabs.harvard.edu/abs/2016NatPh..12..179J} {12, 179}

\bibitem[\protect\citeauthoryear{{Kolotkov}, {Nakariakov}  \&
  {Zavershinskii}}{{Kolotkov} et~al.}{2019}]{2019A&A...628A.133K}
{Kolotkov} D.~Y.,  {Nakariakov} V.~M.,   {Zavershinskii} D.~I.,  2019, \mn@doi
  [\aap] {10.1051/0004-6361/201936072}, \href
  {https://ui.adsabs.harvard.edu/abs/2019A&A...628A.133K} {628, A133}

\bibitem[\protect\citeauthoryear{Kolotkov, Duckenfield  \& Nakariakov}{Kolotkov
  et~al.}{2020}]{Kolotkov_2020}
Kolotkov D.~Y.,  Duckenfield T.~J.,   Nakariakov V.~M.,  2020, \mn@doi [\aap]
  {10.1051/0004-6361/202039095}, 644, A33

\bibitem[\protect\citeauthoryear{Kolotkov, Zavershinskii  \&
  Nakariakov}{Kolotkov et~al.}{2021}]{10.1088/1361-6587/ac36a5}
Kolotkov D.~Y.,  Zavershinskii D.~I.,   Nakariakov V.~M.,  2021, \mn@doi
  [Plasma Physics and Controlled Fusion] {10.1088/1361-6587/ac36a5}, 63, 124008

\bibitem[\protect\citeauthoryear{{Krishna Prasad}, {Raes}, {Van Doorsselaere},
  {Magyar}  \& {Jess}}{{Krishna Prasad} et~al.}{2018}]{2018ApJ...868..149K}
{Krishna Prasad} S.,  {Raes} J.~O.,  {Van Doorsselaere} T.,  {Magyar} N.,
  {Jess} D.~B.,  2018, \mn@doi [\apj] {10.3847/1538-4357/aae9f5}, \href
  {https://ui.adsabs.harvard.edu/abs/2018ApJ...868..149K} {868, 149}

\bibitem[\protect\citeauthoryear{{Li}, {Antolin}, {Guo}, {Kuznetsov}, {Pascoe},
  {Van Doorsselaere}  \& {Vasheghani Farahani}}{{Li}
  et~al.}{2020}]{2020SSRv..216..136L}
{Li} B.,  {Antolin} P.,  {Guo} M.~Z.,  {Kuznetsov} A.~A.,  {Pascoe} D.~J.,
  {Van Doorsselaere} T.,   {Vasheghani Farahani} S.,  2020, \mn@doi [\ssr]
  {10.1007/s11214-020-00761-z}, \href
  {https://ui.adsabs.harvard.edu/abs/2020SSRv..216..136L} {216, 136}

\bibitem[\protect\citeauthoryear{{Marsh}, {De Moortel}  \& {Walsh}}{{Marsh}
  et~al.}{2011}]{Marsh2011}
{Marsh} M.~S.,  {De Moortel} I.,   {Walsh} R.~W.,  2011, \mn@doi [Astrophys J]
  {10.1088/0004-637X/734/2/81}, \href
  {https://ui.adsabs.harvard.edu/abs/2011ApJ...734...81M} {734, 81}

\bibitem[\protect\citeauthoryear{Molevich \& Oraevskii}{Molevich \&
  Oraevskii}{1988}]{Molevich88}
Molevich N.~E.,  Oraevskii A.~N.,  1988, Zh. Eksp. Teor. Fiz, 94, 128

\bibitem[\protect\citeauthoryear{Nakariakov \& Kolotkov}{Nakariakov \&
  Kolotkov}{2020}]{Nakariakov2020}
Nakariakov V.~M.,  Kolotkov D.~Y.,  2020, \mn@doi [Annu Rev of Astron and
  Astrophys] {10.1146/annurev-astro-032320-042940}, 58, 441

\bibitem[\protect\citeauthoryear{{Nakariakov} et~al.,}{{Nakariakov}
  et~al.}{2021}]{2021SSRv..217...73N}
{Nakariakov} V.~M.,  et~al., 2021, \mn@doi [\ssr] {10.1007/s11214-021-00847-2},
  \href {https://ui.adsabs.harvard.edu/abs/2021SSRv..217...73N} {217, 73}

\bibitem[\protect\citeauthoryear{Pascoe, Goddard, Nistic\`{o}, Anfinogentov  \&
  Nakariakov}{Pascoe et~al.}{2016}]{Pascoe2016}
Pascoe D.~J.,  Goddard C.~R.,  Nistic\`{o} G.,  Anfinogentov S.,   Nakariakov
  V.~M.,  2016, \mn@doi [\aap] {10.1051/0004-6361/201628255}, 589, A136

\bibitem[\protect\citeauthoryear{Prasad, Banerjee  \& Doorsselaere}{Prasad
  et~al.}{2014}]{Prasad2014}
Prasad S.~K.,  Banerjee D.,   Doorsselaere T.~V.,  2014, \mn@doi [Astrophys J]
  {10.1088/0004-637x/789/2/118}, 789, 118

\bibitem[\protect\citeauthoryear{{Prasad}, {Srivastava}  \& {Wang}}{{Prasad}
  et~al.}{2021}]{2021SoPh..296..105P}
{Prasad} A.,  {Srivastava} A.~K.,   {Wang} T.,  2021, \mn@doi [\solphys]
  {10.1007/s11207-021-01846-w}, \href
  {https://ui.adsabs.harvard.edu/abs/2021SoPh..296..105P} {296, 105}

\bibitem[\protect\citeauthoryear{{Priest}}{{Priest}}{2014}]{2014masu.book.....P}
{Priest} E.,  2014, {Magnetohydrodynamics of the Sun},
  \mn@doi{10.1017/CBO9781139020732.
}

\bibitem[\protect\citeauthoryear{Roberts}{Roberts}{2019}]{roberts_2019}
Roberts B.,  2019, MHD Waves in the Solar Atmosphere.
Cambridge University Press, \mn@doi{10.1017/9781108613774}

\bibitem[\protect\citeauthoryear{{Rosner}, {Tucker}  \& {Vaiana}}{{Rosner}
  et~al.}{1978a}]{1978ApJ...220..643R}
{Rosner} R.,  {Tucker} W.~H.,   {Vaiana} G.~S.,  1978a, \mn@doi [\apj]
  {10.1086/155949}, \href
  {https://ui.adsabs.harvard.edu/abs/1978ApJ...220..643R} {220, 643}

\bibitem[\protect\citeauthoryear{{Rosner}, {Tucker}  \& {Vaiana}}{{Rosner}
  et~al.}{1978b}]{Rosner1978}
{Rosner} R.,  {Tucker} W.~H.,   {Vaiana} G.~S.,  1978b, \mn@doi [\apj]
  {10.1086/155949}, \href
  {https://ui.adsabs.harvard.edu/abs/1978ApJ...220..643R} {220, 643}

\bibitem[\protect\citeauthoryear{{Van Doorsselaere}, {Wardle}, {Del Zanna},
  {Jansari}, {Verwichte}  \& {Nakariakov}}{{Van Doorsselaere}
  et~al.}{2011}]{2011ApJ...727L..32V}
{Van Doorsselaere} T.,  {Wardle} N.,  {Del Zanna} G.,  {Jansari} K.,
  {Verwichte} E.,   {Nakariakov} V.~M.,  2011, \mn@doi [\apjl]
  {10.1088/2041-8205/727/2/L32}, \href
  {https://ui.adsabs.harvard.edu/abs/2011ApJ...727L..32V} {727, L32}

\bibitem[\protect\citeauthoryear{{Wang}, {Ofman}, {Sun}, {Provornikova}  \&
  {Davila}}{{Wang} et~al.}{2015}]{2015ApJ...811L..13W}
{Wang} T.,  {Ofman} L.,  {Sun} X.,  {Provornikova} E.,   {Davila} J.~M.,  2015,
  \mn@doi [\apjl] {10.1088/2041-8205/811/1/L13}, \href
  {https://ui.adsabs.harvard.edu/abs/2015ApJ...811L..13W} {811, L13}

\bibitem[\protect\citeauthoryear{{Wang}, {Ofman}, {Yuan}, {Reale}, {Kolotkov}
  \& {Srivastava}}{{Wang} et~al.}{2021}]{2021SSRv..217...34W}
{Wang} T.,  {Ofman} L.,  {Yuan} D.,  {Reale} F.,  {Kolotkov} D.~Y.,
  {Srivastava} A.~K.,  2021, \mn@doi [\ssr] {10.1007/s11214-021-00811-0}, \href
  {https://ui.adsabs.harvard.edu/abs/2021SSRv..217...34W} {217, 34}

\bibitem[\protect\citeauthoryear{{Zaitsev} \& {Stepanov}}{{Zaitsev} \&
  {Stepanov}}{1975}]{1975IGAFS..37....3Z}
{Zaitsev} V.~V.,  {Stepanov} A.~V.,  1975, Issledovaniia Geomagnetizmu
  Aeronomii i Fizike Solntsa, \href
  {https://ui.adsabs.harvard.edu/abs/1975IGAFS..37....3Z} {37, 3}

\bibitem[\protect\citeauthoryear{{Zaitsev} \& {Stepanov}}{{Zaitsev} \&
  {Stepanov}}{1982}]{1982SvAL....8..132Z}
{Zaitsev} V.~V.,  {Stepanov} A.~V.,  1982, Sov Astron Lett, \href
  {https://ui.adsabs.harvard.edu/abs/1982SvAL....8..132Z} {8, 132}

\bibitem[\protect\citeauthoryear{Zavershinskii, Kolotkov, Nakariakov, Molevich
  \& Ryashchikov}{Zavershinskii et~al.}{2019}]{Zavershinskii2019}
Zavershinskii D.~I.,  Kolotkov D.~Y.,  Nakariakov V.~M.,  Molevich N.~E.,
  Ryashchikov D.~S.,  2019, \mn@doi [Phys Plasmas] {10.1063/1.5115224}, 26,
  082113

\bibitem[\protect\citeauthoryear{{Zavershinskii}, {Molevich}, {Riashchikov}  \&
  {Belov}}{{Zavershinskii} et~al.}{2020}]{2020PhRvE.101d3204Z}
{Zavershinskii} D.~I.,  {Molevich} N.~E.,  {Riashchikov} D.~S.,   {Belov}
  S.~A.,  2020, \mn@doi [\pre] {10.1103/PhysRevE.101.043204}, \href
  {https://ui.adsabs.harvard.edu/abs/2020PhRvE.101d3204Z} {101, 043204}

\bibitem[\protect\citeauthoryear{{Zavershinskii}, {Kolotkov}, {Riashchikov}  \&
  {Molevich}}{{Zavershinskii} et~al.}{2021}]{2021SoPh..296...96Z}
{Zavershinskii} D.,  {Kolotkov} D.,  {Riashchikov} D.,   {Molevich} N.,  2021,
  \mn@doi [\solphys] {10.1007/s11207-021-01841-1}, \href
  {https://ui.adsabs.harvard.edu/abs/2021SoPh..296...96Z} {296, 96}

\bibitem[\protect\citeauthoryear{Zhugzhda}{Zhugzhda}{1996}]{Zhugzhda96}
Zhugzhda Y.~D.,  1996, \mn@doi [Phys Plasmas] {10.1063/1.871836}, 3, 10

\bibitem[\protect\citeauthoryear{{van der Linden} \& {Goossens}}{{van der
  Linden} \& {Goossens}}{1991}]{1991SoPh..131...79V}
{van der Linden} R.~A.~M.,  {Goossens} M.,  1991, \mn@doi [\solphys]
  {10.1007/BF00151746}, \href
  {https://ui.adsabs.harvard.edu/abs/1991SoPh..131...79V} {131, 79}

\makeatother
\end{thebibliography}

\section*{Appendix}\label{Appen}

Further, we will describe the main mathematical steps taken to derive equations~(\ref{Connect}) -- (\ref{z-comp}). 
Linearizing the system of equations~(\ref{Induction}) -- (\ref{State}) and projecting the obtained equations  onto the coordinate axes will gives us a set of equations shown below:

\begin{equation}
	\pder{{B_1}_x}{t}={B_0}	\pder{{V_1}_x}{z},
	\label{Induction_x1}
\end{equation}
\begin{equation}
	\pder{{B_1}_y}{t}={B_0}	\pder{{V_1}_y}{z},
	\label{Induction_y1}
\end{equation}
\begin{equation}
	\pder{{B_1}_z}{t} = {B_0}  \pder{{V_1}_z}{z} -{B_0} \Theta - {{V_1}_x}	\deriv{B_0}{x} ,
	\label{Induction_z1}
\end{equation}
\begin{equation}
	\grad\cdot\vect{B_1}=0 \,,
	\label{Div_1}
\end{equation}
\begin{equation}
	\rho_0 \pder{{V_1}_x}{t}=-\pder{P_T}{x} + \frac{B_0 }{4 \pi}\pder{{B_1}_x}{z}
	\label{Motion_1x},
\end{equation}
\begin{equation}
	\rho_0 \pder{{V_1}_y}{t}=-\pder{P_T}{y} + \frac{B_0 }{4 \pi}\pder{{B_1}_y}{z},
	\label{Motion_1y}
\end{equation}
\begin{equation}
	\rho_0 \pder{{V_1}_z}{t}=-\pder{P_T}{z} + \frac{B_0 }{4 \pi}\pder{{B_1}_z}{z}+ \frac{{B_1}_x}{4 \pi}\deriv{{B_0}}{x},
	\label{Motion_1z}
\end{equation}
\begin{equation}
	\pder{\rho_1}{t} + \left(\vect{v_1}\cdot \grad\right)	{\rho_0} + \rho_0  \Theta = 0,
	\label{Cont_1}
\end{equation}
\begin{multline}
	\frac{1}{\gamma-1} \left( \pder{{P_1}}{t} + \left(\vect{v_1}\cdot \grad\right){P_0}  \right) - \frac{\gamma}{\gamma-1} 
	\frac{P_0}{\rho_0}\left( \pder{{\rho_1}}{t} + \left(\vect{v_1}\cdot \grad\right){\rho_0}  \right) = \\
	= -\rho_0 \left({Q_{0T}} T_1+{Q_{0\rho}} \rho_1\right),
	\label{Heat_1}
\end{multline}
\begin{equation}
	P_1=\frac{k_\mathrm{B}}{m}\left(\rho_0 T_1 +\rho_1 T_0 \right)\,.
	\label{State_1}
\end{equation}

First, the perturbation of temperature  $T_1$ is expressed in terms of the perturbations of density  $\rho_1$  and pressure $P_1$ using equation \ref{State_1}. By substituting the obtained expression, the temperature perturbation  $T_1$ is excluded from equation  \ref{Heat_1}. The next step is to exclude the density perturbation  $\rho_1$ from the modified equation \ref{Heat_1}. To this end, equation   \ref{Heat_1} is differentiated with respect to time, whereas equation  \ref{Cont_1} is used as a substitution. As a result of several simplifications, the following equation for the perturbations of pressure  $P_1$ and velocity vector $\vect{v_1}$ can be obtained:
\begin{multline}
    \pder{}{t} \left( \pder{{P_1}}{t} + \left(\vect{v_1}\cdot \grad\right){P_0} + \gamma P_0 \Theta \right) = \\
     = -\frac{Q_{0T}}{C_V} \left( \pder{{P_1}}{t} + \frac{k_\mathrm{B}}{m} 
     \frac{\left({Q_{0T}} T_0-{Q_{0\rho}} \rho_0\right)}{{Q_{0T}} } \left( \left(\vect{v_1}\cdot \grad\right){\rho_0} + \rho_0 \Theta \right)      \right). \\
	\label{P_1__V1}
\end{multline}
In terms of characteristic timescales \ref{characteristic_times} and speeds \ref{characteristic_speeds}, this equation can be rewritten as: 
\begin{multline}
	\pder{}{t} \left(\pder{{P_1}}{t} + \left(\vect{v_1}\cdot \grad\right){P_0} + c_\mathrm{S}^2 \rho_0\Theta\right)    = \\
	= -\frac{1}{\tau_{V}} \left( \pder{{P_1}}{t} + 
    c_\mathrm{SQ}^2  \left(\vect{v_1}\cdot \grad\right)\rho_0 +  c_\mathrm{SQ}^2 \rho_0\Theta    \right). 	\label{P_1__V2}
\end{multline}

Our next objective is to obtain equation \ref{Connect} describing the perturbation of total pressure $P_T$. By multiplying equation  \ref{Induction_z1} by $B_0/4 \pi$ and by performing simplification, we obtain the expression shown below:
\begin{equation}
	\pder{}{t} \left(\frac{{B_0 {B_1}_z}}{4 \pi}\right) = \frac{{B_0^2 }}{4 \pi} \left( \pder{{V_1}_z}{z} -\Theta  \right)  -  {{V_1}_x} \frac{{B_0}}{4 \pi}	\deriv{B_0}{x}.
	\label{Induction_z1_mod}
\end{equation}

This equation can have two modifications. The first one is obtained by means of differentiation with respect to time.
For the second modification, equation  \ref{Induction_z1_mod} is multiplied by    $1/\tau_V$. Further, equation \ref{Connect} can be obtained by summing up the two previously derived equations and equation \ref{P_1__V1} and by introducing the Alfven speed $c_A$ and the total pressure perturbation $P_T$.

The equation for the \textit{x}-component of the velocity vector \ref{x-comp} is derived by differentiating equation \ref{Motion_1x} with respect to time and excluding the magnetic field derivative $\partial {B_1}_x / \partial t $ using equation \ref{Induction_x1}. Similarly, equations  \ref{y-comp} and \ref{z-comp} can be derived using equations \ref{Motion_1y}, \ref{Induction_y1} and \ref{Motion_1z}, \ref{Induction_z1}, respectively. 



\bsp	
\label{lastpage}
\end{document}